\documentclass[10pt,conference]{IEEEtran}
\usepackage{cite}
\usepackage{amsmath,amssymb,amsfonts}
\usepackage{algorithmic}
\usepackage{graphicx}
\usepackage{textcomp}
\usepackage{xcolor}
\usepackage[hyphens]{url}

\usepackage{comment}
\usepackage{algorithm}
\usepackage{subfig}
\usepackage{wrapfig}
\usepackage{caption}
\usepackage{xcolor}
\usepackage{soul}

\def\BibTeX{{\rm B\kern-.05em{\sc i\kern-.025em b}\kern-.08em
    T\kern-.1667em\lower.7ex\hbox{E}\kern-.125emX}}

\newcommand{\hpcasubmissionnumber}{302}

\title{ACIC: Admission-Controlled Instruction Cache} 

\author{
    \IEEEauthorblockN{Yunjin Wang\IEEEauthorrefmark{1}, Chia-Hao Chang\IEEEauthorrefmark{1}, Anand Sivasubramaniam\IEEEauthorrefmark{1}, Niranjan Soundararajan\IEEEauthorrefmark{2}{\textsection}}
    \IEEEauthorblockA{\IEEEauthorrefmark{1}The Pennsylvania State University\\
    \IEEEauthorrefmark{2}Intel Labs, India \\
    }
    \IEEEauthorblockA{
    \IEEEauthorrefmark{1}\{yjw5279, cuc1057, axs53\}@psu.edu \\
    \IEEEauthorrefmark{2}\{niranjan.k.soundararajan@intel.com}
}

\begin{document}
\maketitle

\begingroup\renewcommand\thefootnote{\textsection}
\footnotetext{This work was done while the author was at Intel Labs.}
\endgroup

\thispagestyle{plain}
\pagestyle{plain}

\begin{abstract}
The front end bottleneck in datacenter workloads has come under increased scrutiny, with the growing code footprint, involvement of numerous libraries and OS services, and the unpredictability in the instruction stream. Our examination of these workloads points to burstiness in accesses to instruction blocks, which has also been observed in data accesses~\cite{cache-bursts}. Such burstiness is largely due to spatial and short-duration temporal localities, that LRU fails to recognize and optimize for, when a single cache caters to both forms of locality. Instead, we incorporate a small i-Filter as in previous works~\cite{filter-cache,small-icache} to separate spatial from temporal accesses. However, a simple separation does not suffice, and we additionally need to predict whether the block will continue to have temporal locality, after the burst of spatial locality. This combination of i-Filter and temporal locality predictor constitutes our Admission-Controlled Instruction Cache (ACIC). ACIC outperforms a number of state-of-the-art pollution reduction techniques (replacement algorithms, bypassing mechanisms, victim caches), providing 1.0223 speedup on the average over a baseline LRU based conventional i-cache (bridging over half of the gap between LRU and OPT)
across several datacenter workloads.
\end{abstract}


\section{Introduction}

The front-end stalls in datacenter applications have come under much scrutiny in recent years. These applications have complex and deep software stacks, 
executing millions of instructions even for a single user query~\cite{memory-hierachy-for-web-search}. The consequent unpredictability in their control flow, involvement of numerous software layers (libraries and OS) beyond the application, and the resulting large code footprint have been noted to cause higher instruction cache (referred henceforth as i-cache) misses compared to the more conventional scientific and desktop workloads, like SPEC~\cite{spec2017}. 
One can attempt to prefetch instruction blocks, and/or predict branches, based on anticipated control flow, which can help reduce i-cache misses and branch mispredictions. Another important angle for attacking this problem is by being more discretionary in what to bring and retain (replacement algorithm) within the precious and limited i-cache space. Taking the latter approach, this paper draws insights from prior work~\cite{cache-bursts} that proposes dead block predictors based on bursty accesses to a cache block, and finds that such bursty accesses are also widespread in datacenter workloads, due to distinct spatial and temporal localities in the instruction stream, that are often not well serviced by the conventional LRU replacement algorithm in a single i-cache.
Instead, the paper adds a separate and small (16-entry) buffer, whose similar/variant forms have been proposed in prior works~\cite{filter-cache, small-icache}, to meet spatial localities, and implements an admission control mechanism to determine whether the block will continue to have temporal locality to justify bringing it into the i-cache. This Admission-Controlled Instruction Cache (ACIC) delivers 1.0223 speedup on the average across 10 datacenter applications 18.14\% reduction in i-cache misses), buying back 55.85\% of the performance loss of the baseline LRU over the oracle-based optimal (OPT) replacement algorithm that is theoretically possible.

While one may question the motivation for reducing i-cache misses given that it typically results in single-digit percentage speedups, as pointed out in~\cite{softsku}, achieving even these single-digit percentage speedups is important to provide significant performance-per-watt benefits in these workloads.
Consequently, there have been
numerous studies looking to reduce i-cache misses~\cite{DBMS-where-does-time-go, reactive-nuca, profile-warehouse, TIFS, PIF, pTask, RDIP}. These techniques can be categorized into: (i) prefetching mechanisms (whether purely in hardware~\cite{TIFS, PIF, Confluence, Boomerang, Shotgun, SHIFT, divide-and-conquer, EIP} or through profile-guided~\cite{AsmDB, I-SPY} and compilation~\cite{call-chain-prefetch, cooperative-prefetch} techniques); (ii) code re-layout optimizations for better instruction locality~\cite{Ispike, code-layout-optimization, lightweight-feedback-directed, AutoFDO, optimal-function-placement, BOLT}; and
(iii) better management of i-cache space by controlling what should be brought in/replaced (e.g. GHRP~\cite{GHRP}, Ripple~\cite{Ripple}). This work falls in category (iii), and we will show that it can complement some of the recently proposed prefetching techniques of category (i) as well.

Conventionally, the L1 instruction caches (i-cache) have been considered with a relatively small (4 or 8) associativity and a LRU-based replacement algorithm within a set. To a large extent, this structure has served its purpose fairly well for the much smaller code footprints. However, with the larger footprints of datacenter workloads, it is not clear whether the traditional LRU would work as well. This has also been a reason for some of the recent efforts such as GHRP~\cite{GHRP} and Ripple~\cite{Ripple} which have tried to improve the replacement mechanism by predicting reuse distances and identifying problematic program fragments. LRU relies on the recent past to predict the future, and may not be suitable for some of the bursty scenarios that we observe in emerging datacenter applications~\cite{cloudsuite, oltpbench, renaissance}.


In this paper, we draw insight from prior work~\cite{cache-bursts} which uses cache burst history, instead of cache access history, to predict dead blocks. Unsurprisingly, such bursty accesses are also common in the instruction stream of these 
datacenter workloads: (a) {\em a block that is referenced, continues to experience considerable spatial and short-term temporal locality, i.e. a burst}. This is fairly intuitive since successive instructions of the stream would fall in the same block. There is also short-term temporal locality due to locality in the recent branch targets, as pointed out in~\cite{PIF}.
(b) {\em after this burst, it is not very clear whether the block is more important than another block which may already be present in the i-cache}. It may happen that after this burst, the block may not be needed for a long time (its reuse distance is much longer) that it better not be brought into the i-cache to result in pollution. 
A single i-cache with LRU-based replacement policy, would not differentiate between the accesses within a burst and those between bursts. This is a reason why streaming buffers~\cite{victim-cache, stream-buffer-as-secondary-cache, filter-cache,small-icache}, separate caches for the two forms of locality~\cite{temporal-and-spatial-locality-caches, spatial-temporal-locality-aware-cache, multiple-caching-strategies}, and/or cache bypassing schemes~\cite{runtime_cache_bypassing}, have been proposed to handle the two localities differently. 

Based on this observation, we add an i-Filter (whose similar forms have been proposed and studied in prior works~\cite{filter-cache, small-icache}), which is a 16-slot buffer for instruction blocks to handle the spatial and short-term temporal accesses. 
However, when this buffer becomes full, the victim cannot be simply evicted (then i-cache serves no purpose) or simply inserted into i-cache (which can cause pollution). Instead, we need a prediction mechanism to determine whether its reuse distance (i.e. to the next burst) is shorter than a block already in the i-cache that it will replace. If so, and only then, should we bring it into i-cache. This overall mechanism, termed Admission-Controlled Instruction Cache (ACIC), provides the necessary spatio-temporal separation to differentially meet the intra-burst and inter-burst accesses that LRU is not tuned for.


This paper makes the following contributions to reduce i-cache misses in datacenter workloads:

\begin{itemize}
    \item We show that accesses to an instruction block are bursty, similar to the observation for data accesses in~\cite{cache-bursts},
    with a large number of spatial and near-term temporal accesses, rather than spread out over the execution. 
    \item At the same time, one cannot ignore the separation between the bursts. LRU, in such cases, presumes the block will continue to be needed soon, thus reaching a wrong decision in bringing it into i-cache. 
    \item Further, we cannot throw away the block after its burst is done either. With several blocks being needed, we need to be very discretionary about what to retain and what to throw away from the i-cache.
    \item We present the design of ACIC which provisions (i) a 16-entry i-Filter to temporarily hold incoming blocks for accesses during a burst, and (ii) a prediction mechanism to determine whether its reuse distance after the burst is shorter than a contender block already in its i-cache set, and filtering it out otherwise.
    \item Using a number of datacenter applications we show that ACIC reduces i-cache misses by 18.14\% (with a standard fetch-directed prefetcher~\cite{FDP}). This results in a 1.0223 speedup, bridging over half of the gap between conventional LRU and OPT (which is not implementable). The hardware takes 2.67KB (around 2/3rd of some other recent proposals) space and  saves 0.63\% chip energy over the baseline system.
    \item We also show that ACIC provides better performance than other recently proposed cache replacement policies (GHRP~\cite{GHRP}, SRRIP~\cite{SRRIP}, SHiP~\cite{SHiP}, Hawkeye~\cite{Hawkeye}/Harmony~\cite{Harmony}), cache bypassing policies (DSB~\cite{DSB} and OBM~\cite{OBM} which were initially proposed for d-caches), and alternate strategies such as victim caches (VVC~\cite{VVC} which were earlier proposed for d-caches).  
\end{itemize}

\vspace{0.6cm}


\section{Motivation}\label{motivation}

\begin{figure*}[!t]
    \centering
 \subfloat[\small Distribution of reuse distances
\label{reuse_dist_distribution}]{
  \resizebox{0.45\linewidth}{!}{\includegraphics{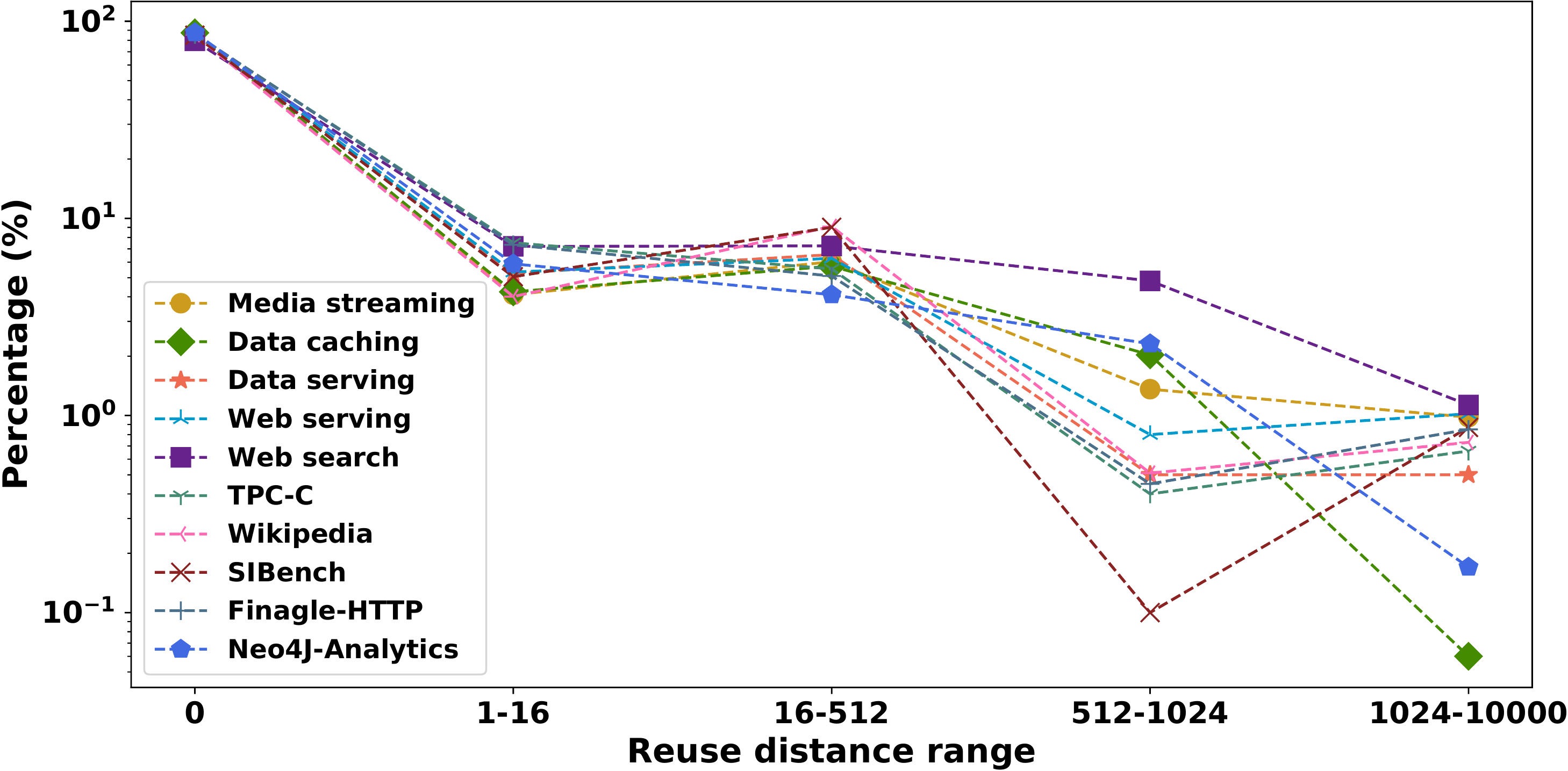}}}
  \subfloat[\small Markov chain of reuse distances in Media Streaming
 \label{markov_chain}]{
  \resizebox{0.45\linewidth}{!}{\includegraphics{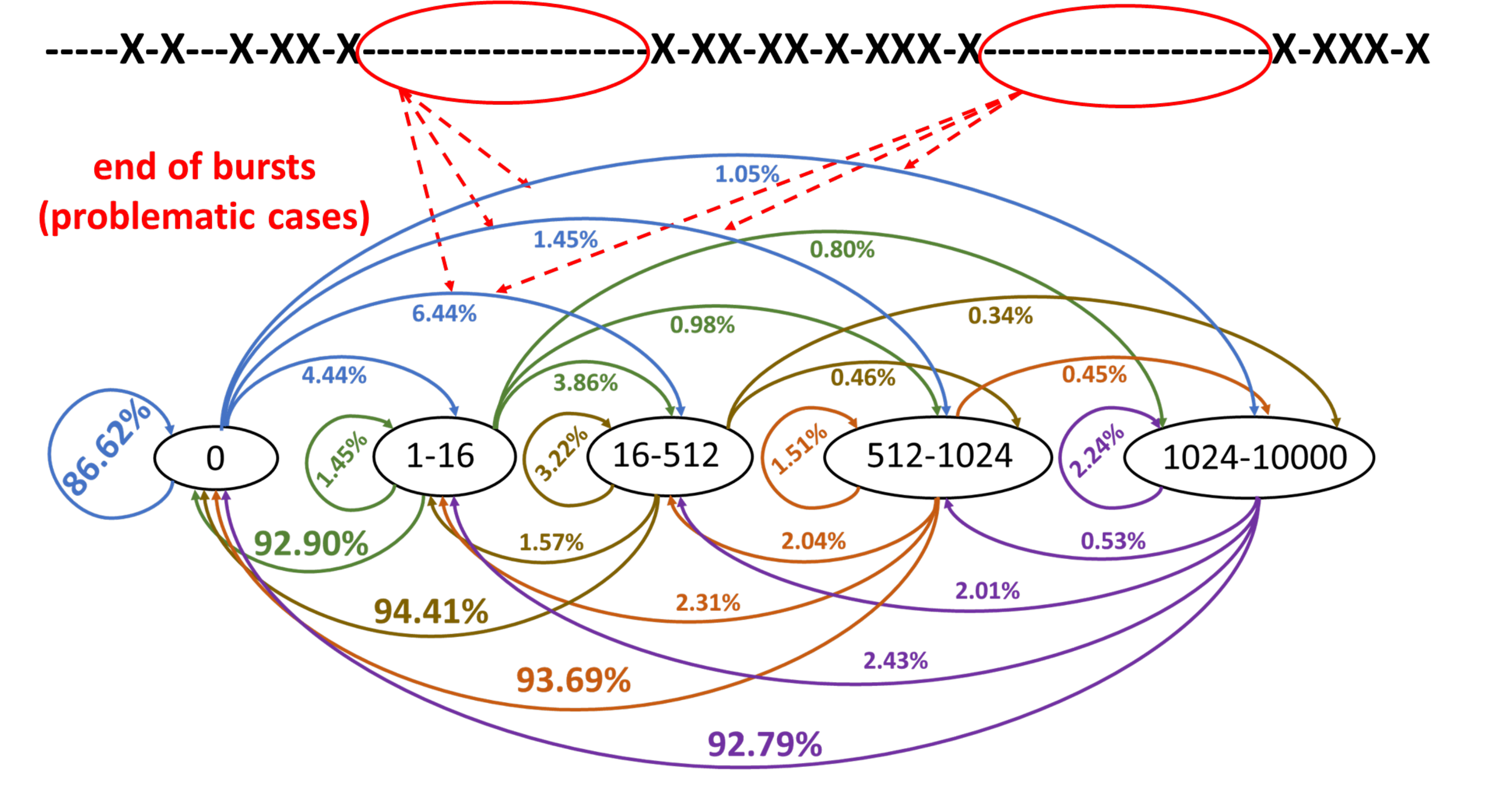}}}
  \caption{Reuse distance analysis. Shows the need for separating spatio-temporal localities.}
  \label{spatio-temporal} 
\end{figure*}

\noindent \underline{\em Need for Spatio-Temporal Separation:} 
Inspired by prior work\cite{cache-bursts} which observes bursty accesses to data blocks and proposes better cache management policies based on such burstiness, we explore similar optimizations for the instruction stream in datacenter workloads~\cite{cloudsuite, oltpbench, renaissance}. We first study the reuse distances of instruction blocks in server workloads, previously identified to have a front-end bottleneck\footnote{In this work, the term \emph{reuse distance} is defined similar to \emph{stack distance} of LRU, i.e. the number of unique instruction cache blocks accessed between two successive accesses to the same instruction block.}.  Figure~\ref{reuse_dist_distribution} plots the distribution of reuse distances between current and previous accesses to the same instruction block. They are histogrammed into buckets on the x-axis for interesting reuse distance ranges (0 implies spatial locality to the same block, 1-16 for very short-term temporal locality, 16-512 captures the size of an i-cache, 512-1024 for distances just out-of-reach of i-cache, and much larger reuse distances (1024-10000). 

In around 85\% of the cases, an instruction block continues to be re-accessed, indicating strong spatial locality (due to successive instructions lying in the same block).
This is typically followed by the [1-16] bucket, indicating high short-term temporal locality. While the log-scale in y-axis does indicate a significant drop as we move beyond reuse-distances that can be captured by today's i-cache sizes (until 512), there is a non-negligible fraction (up to 6\%) that falls beyond the reach of i-cache. While one would think these misses could be ignored, such misses can still amount to as much as 8.23\% loss in speedup in some applications (due to the cost of servicing a miss). This is the region that this paper sets out to optimize. Incidentally, note that even an oracle-based Belady's OPT replacement algorithm~\cite{belady}, tries to optimize for this region.

To further illustrate the spatial locality,
in Figure~\ref{markov_chain}, we show the correlation between successive reuse distances as a Markov Chain in Media Streaming. Each state represents the range of reuse distances, and the transition from one to another indicates the probability of the next reuse distance from the current reuse distance. 
Again, the self-transitions/transitions into the smallest reuse distance states (particularly 0) dominate. This diagram indicates the ``burstiness'' of accesses to instruction blocks, i.e. once a block is referenced, it continues to get referenced for a while (largely due to spatial locality) as is illustrated in the diagram on top of Figure~\ref{markov_chain}. After this burst, its reuse distance can become long again. When this happens, LRU-like schemes which look at the past (rather than the future), try to retain  rather than evict such blocks. 

These two sets of results point to the need to optimize for both spatial and temporal localities of instruction blocks. Currently, both these forms of locality are fulfilled by the single i-cache. However, as has been well known~\cite{LRFU, ARC}, a single cache (with LRU replacement) is not well suited to meet both these forms of locality simultaneously. This is also one of the reasons why streaming buffers~\cite{victim-cache, stream-buffer-as-secondary-cache, filter-cache, small-icache}, separate caches for the two forms of locality~\cite{temporal-and-spatial-locality-caches, spatial-temporal-locality-aware-cache, multiple-caching-strategies}, and/or cache bypassing schemes~\cite{runtime_cache_bypassing}, have been proposed to handle the two localities differently.

It is very likely that a block has already been evicted from and misses in i-cache when it is re-accessed again after the longer reuse distance from its last burst. Our proposed spatio-temporal separation for i-cache aims to eliminate such misses if the block turns out to be useful enough to be retained in i-cache after its burst. While a prefetcher can also try to reduce such misses by predictions, we should note (and will show experimentally) that the two techniques - ACIC and prefetching - are complementary. In fact, as we will show, ACIC can reduce such misses even when a state-of-the-art prefetcher (e.g. \cite{FDP, EIP}) falls short. This goes to show that there is headroom for replacement policies and bypassing policies beyond what prefetchers can provide to further improve i-cache performance.

\begin{figure}[h!]
\begin{center}
  \resizebox{0.80\linewidth}{!}{\includegraphics{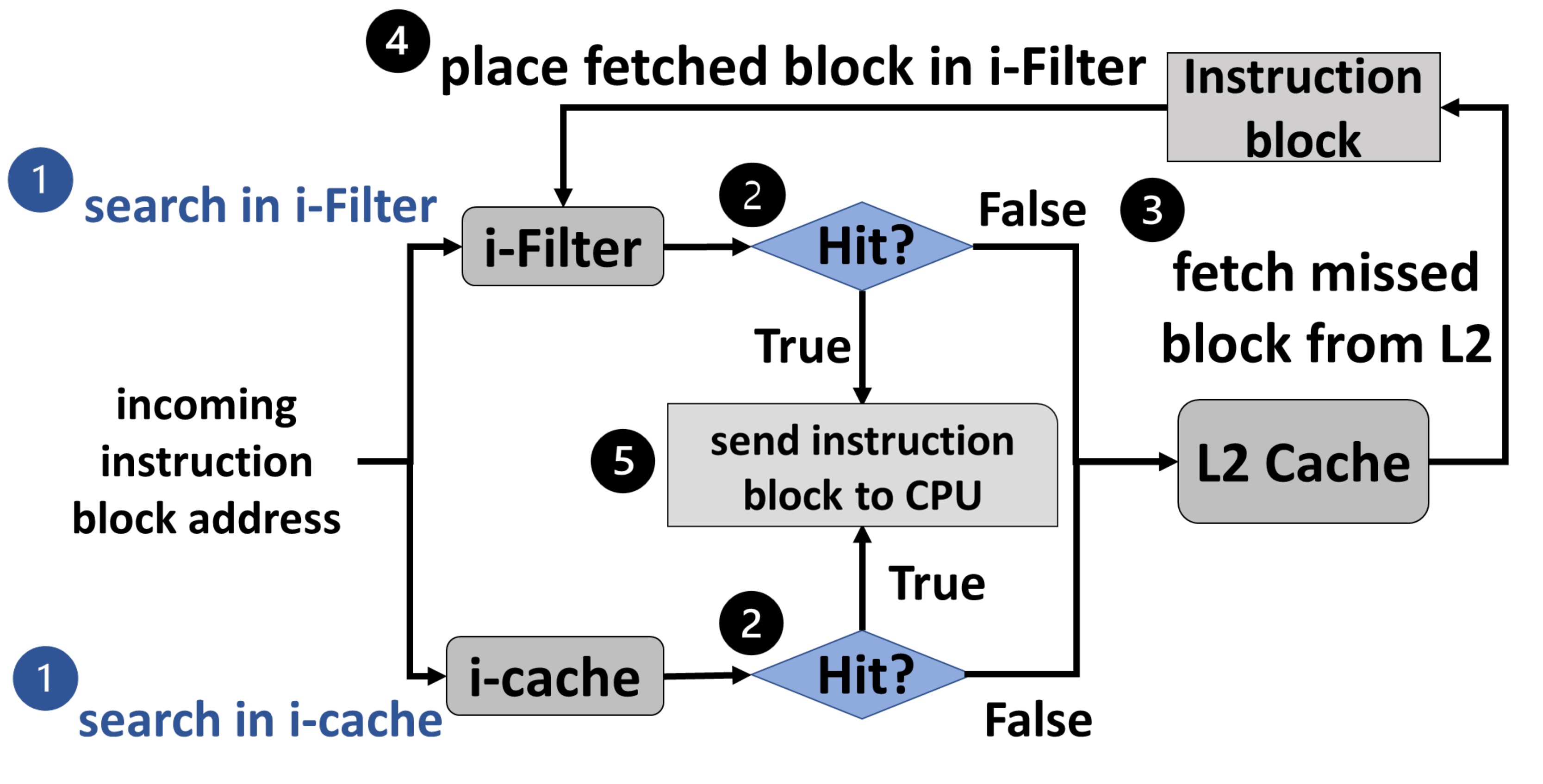}}
  \caption{Instruction block access datapath with i-Filter}
  \label{lookup_phase}
\end{center}
\end{figure}

\noindent \underline{\em Need for further admission control:}
To provide spatio-temporal separation,
we add a 16-slot fully associative buffer, called \emph{i-Filter}, residing next to i-cache, similar to that in~\cite{filter-cache, small-icache}. As Figure~\ref{lookup_phase} shows, upon a fetch, the requested address is searched concurrently in both i-Filter and i-cache. If found in either, the instruction block is sent to CPU, and is considered a hit. Otherwise, the missed block is fetched from deeper in the memory hierarchy and is then placed in i-Filter only. If i-Filter is full, the LRU block in i-Filter is evicted in order to make space. Victim blocks that are evicted from i-Filter are always inserted into i-cache for now.
This i-Filter can fulfill much of the spatial locality. Only when evicted from this structure and inserted into i-cache, will it be subject more to the temporal access patterns for subsequent replacement. 

\begin{figure*}[!t]
\centering
  \subfloat[\small Always inserting i-Filter victims to i-cache provides 1.0057 geomean speedup over baseline. Bypassing with access count comparison provides 1.0102 geomean speedup. OPT replacement policy provides 1.0398 geomean speedup.
   \label{ibuff_only_vs_OPT}]{
        \includegraphics[width=0.485\linewidth]{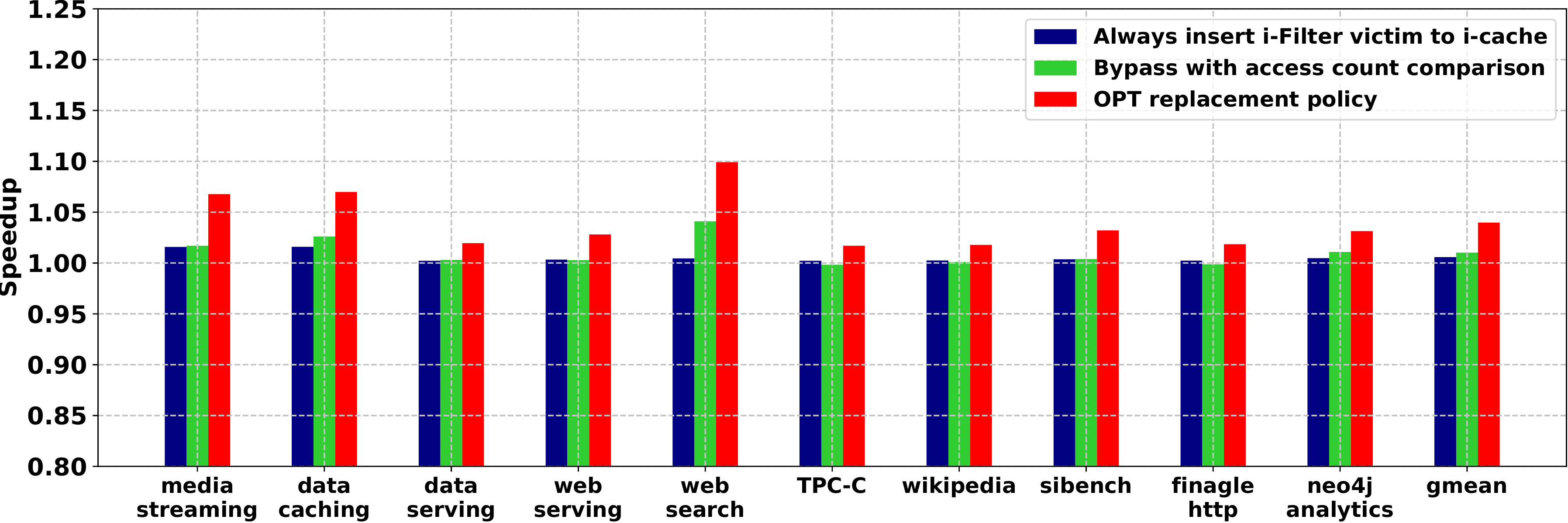}}
\hfill
\hspace{0.1cm}
  \subfloat[\small Reuse distance of incoming block (from i-Filter into i-cache) $-$ reuse distance of outgoing block (selected from the corresponding set of i-cache using OPT) in Media Streaming. In 38.38\% of the cases, the incoming block has a larger reuse distance, showing that the last access of the burst (spatial locality) should not be used for projecting future temporal reuse.
  \label{reuse_dist_difference}]{
        \includegraphics[width=0.48\linewidth]{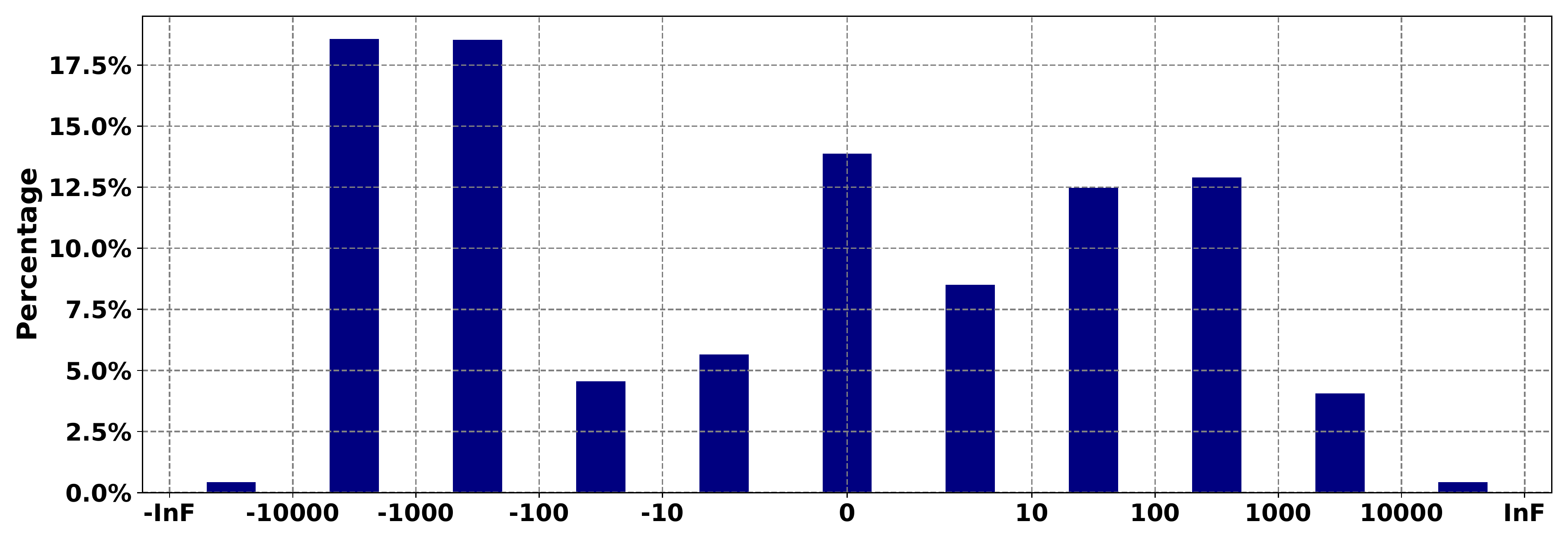}}
    \\
  \caption{Media streaming: Need for additional filtering of blocks from entering L1i after the current burst of accesses in i-Filter}
  \label{admission_control_necessary}
\end{figure*}

We evaluate such a spatio-temporal separation using this i-Filter + i-cache design for 10 widely used datacenter applications listed in Table~\ref{table:benchmarks}. The experimental setup and simulation parameters are described in Section~\ref{evaluation}. Figure~\ref{ibuff_only_vs_OPT} shows the speedup comparison between this scheme and the OPT replacement policy (for i-cache). While the OPT replacement policy provides a 1.0398 speedup over the LRU baseline on average, the i-Filter$+$i-cache scheme provides a measely 1.0057 average speedup, i.e. the spatio-temporal separation with i-Filter is not very effective. The reason behind this gap is that some of the i-Filter victims can cause i-cache pollution, and thus should not be placed in i-cache. 

This happens when the current burst of accesses for a block is done, and the next reuse distance is much larger. At this point, it is not clear whether this reuse distance is larger or smaller than the block in i-cache that it may evict in the corresponding set. If we could find out that this was larger (through oracle knowledge), we would not be inserting it into i-cache.
In Figure~\ref{reuse_dist_difference}, we plot the subsequent reuse distance of the block being inserted from the i-Filter into i-cache, and substract this from the reuse distance of the block that is being evicted from the corresponding i-cache set (selected using OPT). Ideally, this newly inserted block into i-cache should have its next access earlier than the next access for the victim evicted from i-cache, i.e. the percentages of $x$ values greater than 0 should be 0. However, as we can see, nearly 40\% of the time, we are making a wrong decision in moving the block from i-Filter to i-cache.

This suggests that a simple separation of spatial (using i-Filter) and temporal (using i-cache) localities with different structures will not suffice. We additionally {\em need an admission control mechanism to determine whether the block evicted from i-Filter should replace some other block in the corresponding set of i-cache, or whether it should be thrown away,}
motivating our ACIC admission-controlled i-cache. 
We can also use Figure~\ref{reuse_dist_distribution} as an indicator of when such admission control would really matter for an application. In applications such as \emph{Web search}, \emph{Neo4J-analytics}, \emph{Data caching}, and \emph{Media streaming}, we see the intermediary range (512-1024, which is just beyond i-cache's reach), more prominent than even larger reuse distances. These are the cases when comparing reuse distances become more important, as opposed to applications such as \emph{TPC-C} and \emph{Wikipedia} which have much larger reuse distances. 

\section{Solution}
While i-Filter can provide some amount of separation of spatial vs. temporal locality, the key challenge is on what we should do when we have to evict a block from i-Filter (due to capacity):
{\em should we simply throw it away or should we insert it into the corresponding set of i-cache} (hoping that it will be more useful than an existing block in that set)? 
Throwing it out from the i-Filter  blindly implies that we are not leveraging i-cache's capacity. 
On the other hand, while inserting every block evicted from the i-Filter into i-cache (and evicting the LRU candidate from that set) does provide 1.0057 speedup over the baseline on the average across a number of applications as shown in Figure \ref{ibuff_only_vs_OPT}, it falls significantly short of the potential that an OPT replacement algorithm would provide without an i-Filter at all. 
This suggests that we need a more sophisticated mechanism to decide whether to insert the block from i-Filter into the i-cache upon its eviction from the former structure, which we explore 
in this section.
This decision depends on whether the victim from the i-Filter has a reuse distance smaller than that of the block in the corresponding set of i-cache that it will replace, which we will henceforth refer to as the {\em contender block}. 

This issue has been explored to some extent for data caches in a prior cache bypassing work in ~\cite{runtime_cache_bypassing} where access counters of the two are compared, and whichever is larger is retained in the cache and the other is evicted. 
We can apply the same mechanism to our i-Filter victim, and insert it into i-cache if its access count exceeds that of the i-cache contender.
However, unlike the data cache study of \cite{runtime_cache_bypassing}, as Figure \ref{ibuff_only_vs_OPT} shows, this mechanism does not work well for instruction blocks.

Instead of an access count history, we would like to observe patterns in this history of the relative utility of the i-Filter victim vs. the contender in i-cache, to determine
what we should retain in i-cache. Such history needs to be long enough to effectively capture the complex pattern of past reuse distance comparison results of an i-Filter victim and its i-cache contender. The well studied two-level branch predictor~\cite{two_level_branch_predict} is good at maintaining long histories in an abridged form - predicts based on not only the history of the last $n$ branches, but also the record of the last $m$ occurrences of a certain history. It effectively compresses a long history to a smaller representation. Hence, we leverage the two-level branch predictor~\cite{two_level_branch_predict} to develop our filtering mechanism as to which (the i-Filter victim vs. the i-cache contender) 
to admit/retain in i-cache.

\subsection{Two-level Predictor-based Admission Control}
\begin{figure}[h!]
\begin{center}
  \vspace{-0.3cm}
  \resizebox{0.80\linewidth}{!}{\includegraphics{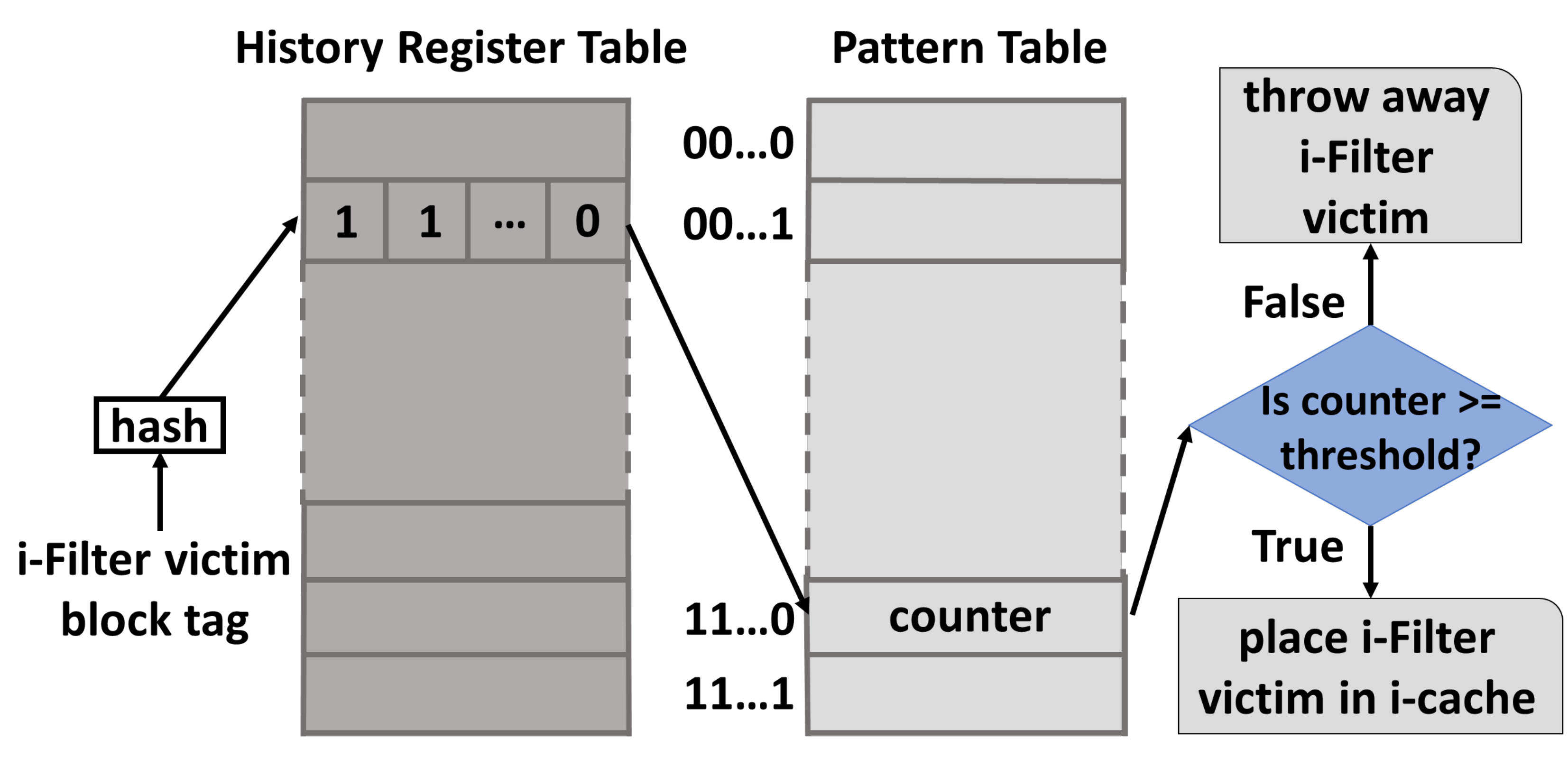}}
  \caption{Two-level i-cache admission predictor}
  \label{predictor_design} 
\end{center}
\end{figure}

Figure~\ref{predictor_design} illustrates our predictor for making this decision.
Similar to the two-level branch predictor~\cite{two_level_branch_predict}, our two-level i-cache admission predictor is comprised of two major data structures: a comparison History Register Table (HRT) and a Pattern Table (PT). 
The tag of an i-Filter victim block is first hashed to index HRT. Each HRT entry is a history register that shifts left with bits which represent the last few comparison results of an i-Filter victim block and its i-cache contender block. If the former is re-accessed sooner in the future\footnote{Section ~\ref{ssec:introduce_CSHR} will discuss how to track future accesses.} than the latter, the history register that the i-Filter victim block is mapped to is shifted left and a 1 is inserted into the least significant bit (LSB).  Else, it is shifted left and a 0 is inserted into the LSB.
PT is indexed by the content of history register. Each PT entry contains a saturating counter that is incremented each time that the i-Filter victim is re-accessed sooner than the i-cache contender block, and is decremented otherwise. A simple threshold is then used to determine whether the i-Filter victim is to be inserted (in place of the contender) in i-cache, or simply thrown away. 

\subsection{Comparing next accesses of i-Filter victim and i-cache contender}\label{ssec:introduce_CSHR}
\begin{figure}[h!]
\begin{center}
  \resizebox{0.85\linewidth}{!}{\includegraphics{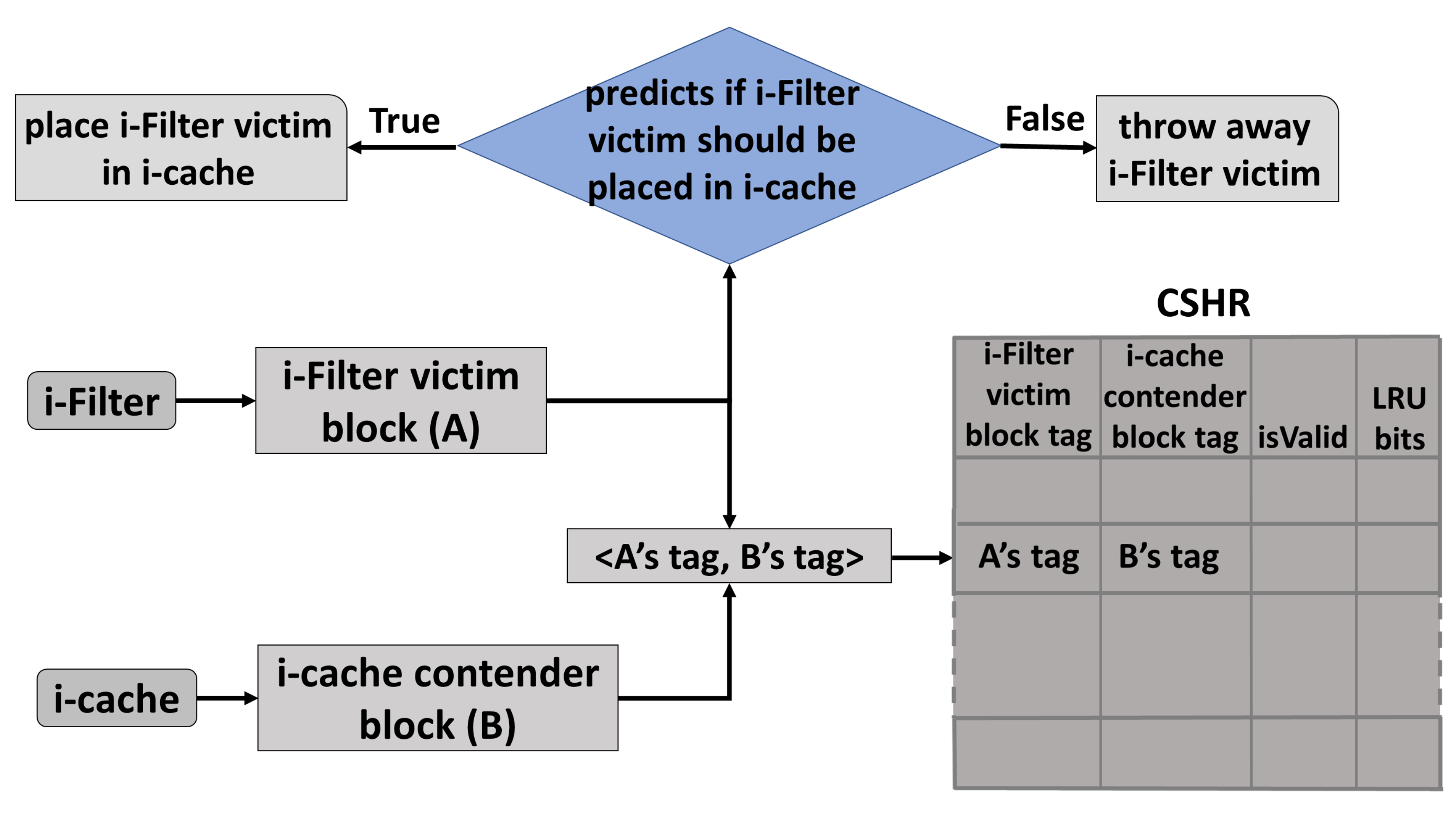}}
  \caption{Insert a new CSHR entry upon i-Filter victim prediction}
  \label{fill_to_CSHR}    
\end{center}
\end{figure}

\vspace{-0.1cm}
To update HRT entries and the counters in PT, we must find out whether an i-Filter victim block will be re-accessed in the nearer future (shorter reuse distance) than its i-cache contender block. Inspired by the design of MSHR (Missing Status Holding Registers) that tracks outstanding misses~\cite{MSHR}, we use a similar structure called CSHR (Comparison Status Holding Registers) to keep track of pairs of i-Filter victim blocks and their i-cache contenders whose comparison results are not yet resolved as shown in Figure~\ref{fill_to_CSHR}. When an i-Filter victim block is being evicted, its tag and the tag of its i-cache contender are inserted into a CSHR entry. As a result of this, the LRU entry in CSHR may need to be evicted since it has a finite size (discussed in Section~\ref{sssec:CSHR_size}). 

When the pipeline front-end issues fetch requests to i-cache in order, the tag of the instruction block being fetched is searched in CSHR. If it matches the i-Filter victim block tag field in a CSHR entry, it means that the i-Filter victim block in the entry is re-accessed sooner than its i-cache contender block. Therefore, the HRT entry is left shifted with a 1 and the corresponding counter of this pattern in PT is incremented. 
On the other hand, if the tag of the instruction block being fetched matches the i-cache contender block tag in a CSHR entry, the PT counter of the HRT entry is decremented with the HRT entry left shifted with a 0.  In either case, as long as the comparison result of a CSHR entry is resolved, this CSHR entry is marked as invalid and can be reused. 

\subsection{Discussion about CSHR}
\subsubsection{Storage overhead}\label{sssec:CSHR_size}
\begin{figure}[h!]
\begin{center}
  \includegraphics[width=\linewidth]{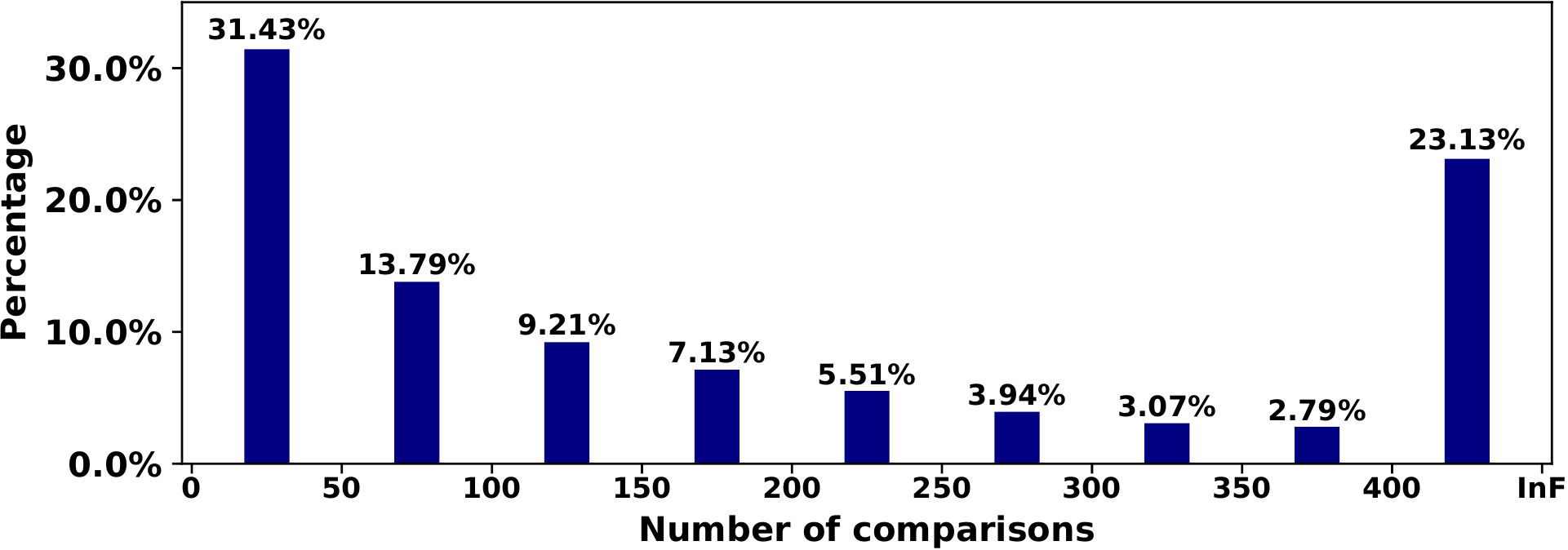}
  \caption{Distribution of comparisons during the lifetime of CSHR entries in Data Caching}
  \label{comparison_distribution}    
\end{center}
\end{figure}

A large CSHR that contains many entries can track more outstanding i-Filter and i-cache block pairs, but it incurs higher storage overhead. On the other hand, with a small CSHR, entries are more likely to be evicted before comparisons are resolved, 
leading to less accurate predictions.
To study the balance between these two factors, we plot Figure~\ref{comparison_distribution}
which shows the incremental percentage of comparisons performed as we increase CSHR entries for a fully associative CSHR design. 
Although around 23\% of comparisons require a very large number of CSHR entries, we find that nearly 70\% of the comparisons get done with just 256 CSHR entries.
Consequently we simply use a 256 entry CSHR, and for those entries which get evicted before being resolved, we give the benefit of doubt to the i-Filter victim as if it was re-accessed earlier than its i-cache contender.

To further reduce storage overhead, instead of full tags of i-Filter victim blocks and i-cache contenders, partial tags are stored in each CSHR entry. When HRT is accessed for either prediction or predictor updates, partial tag, rather than the full block address, of the i-Filter victim block is hashed to index HRT. 
We will show later in Section~\ref{ssec:ACIC_insights} that a 12-bit tag suffices for our needs. Consequently
the CSHR totally takes 256$\times$(2$\times$12-bit tags + 1-bit valid + 5-bit LRU) = 0.9375 KB of space.

\subsubsection{Access cycles}\label{sssec:access_cycles}
\begin{figure}[h!]
\begin{center}
  \includegraphics[width=0.95\linewidth]{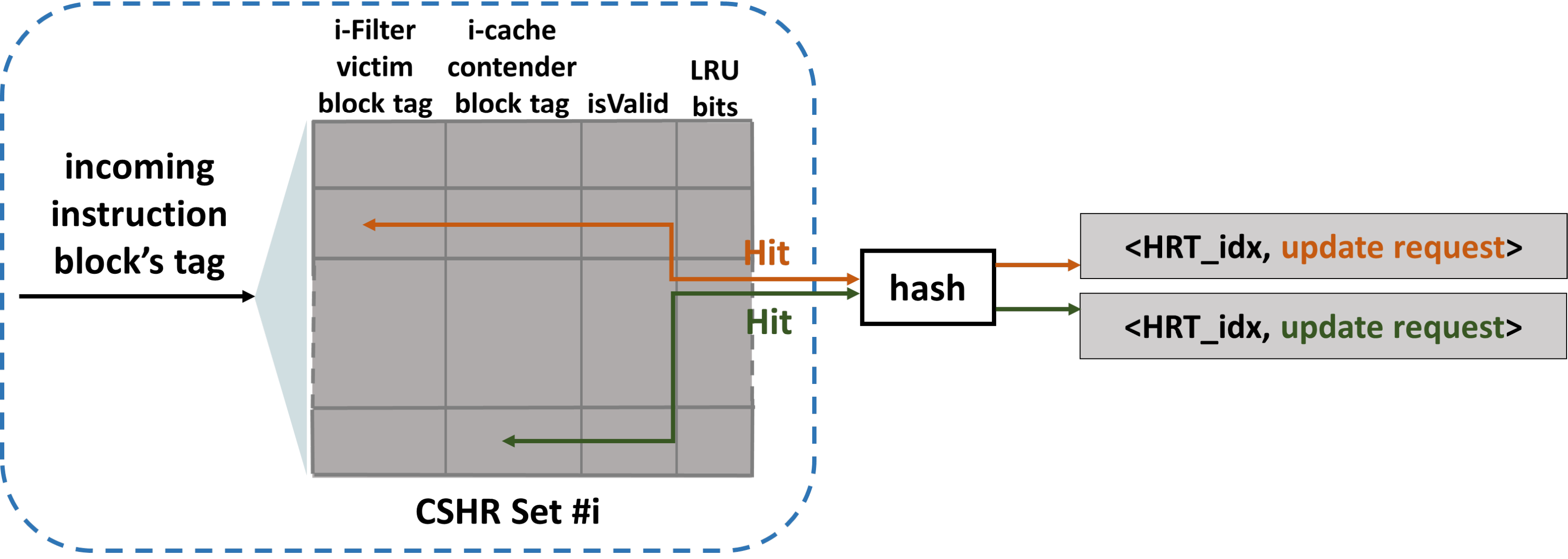}
  \caption{Search incoming block in set-associative CSHR}
  \label{set_assoc_CSHR}    
\end{center}
\end{figure}

Since fetching an instruction block from i-cache can proceed in parallel with searching the partial tag of the instruction block in CSHR, accessing CSHR and updating predictor tables are not in the critical path to accessing i-cache. However, it is not practical to finish searching all the 256 entries in CSHR within one CPU cycle. To solve this problem, we adopt a set associative design for CSHR, in which the 256 entries are divided into $k$ sets. Since the i-Filter block address and the i-cache contender block address in a CSHR entry are always mapped to the same i-cache set, we use the $m$ most significant bits in the i-cache set index to find out to which CSHR set this pair should be inserted. When the instruction block being fetched needs to be searched in CSHR, the $m$ most significant bits in its i-cache set index are used to index the CSHR set and parallel search is done within that set. 
We find that a $k$ value of 8 and a $m$ value of 3 to index the 8 CSHR sets, works quite well for our needs. Each CSHR set adopts LRU as the replacement policy. Figure~\ref{set_assoc_CSHR} shows how the instruction block being fetched is simultaneously searched in the set-associative CSHR and how the predictor tables are updated when CSHR entries are matched. 

\begin{figure}[h!]
\begin{center}
  \resizebox{0.80\linewidth}{!}{\includegraphics{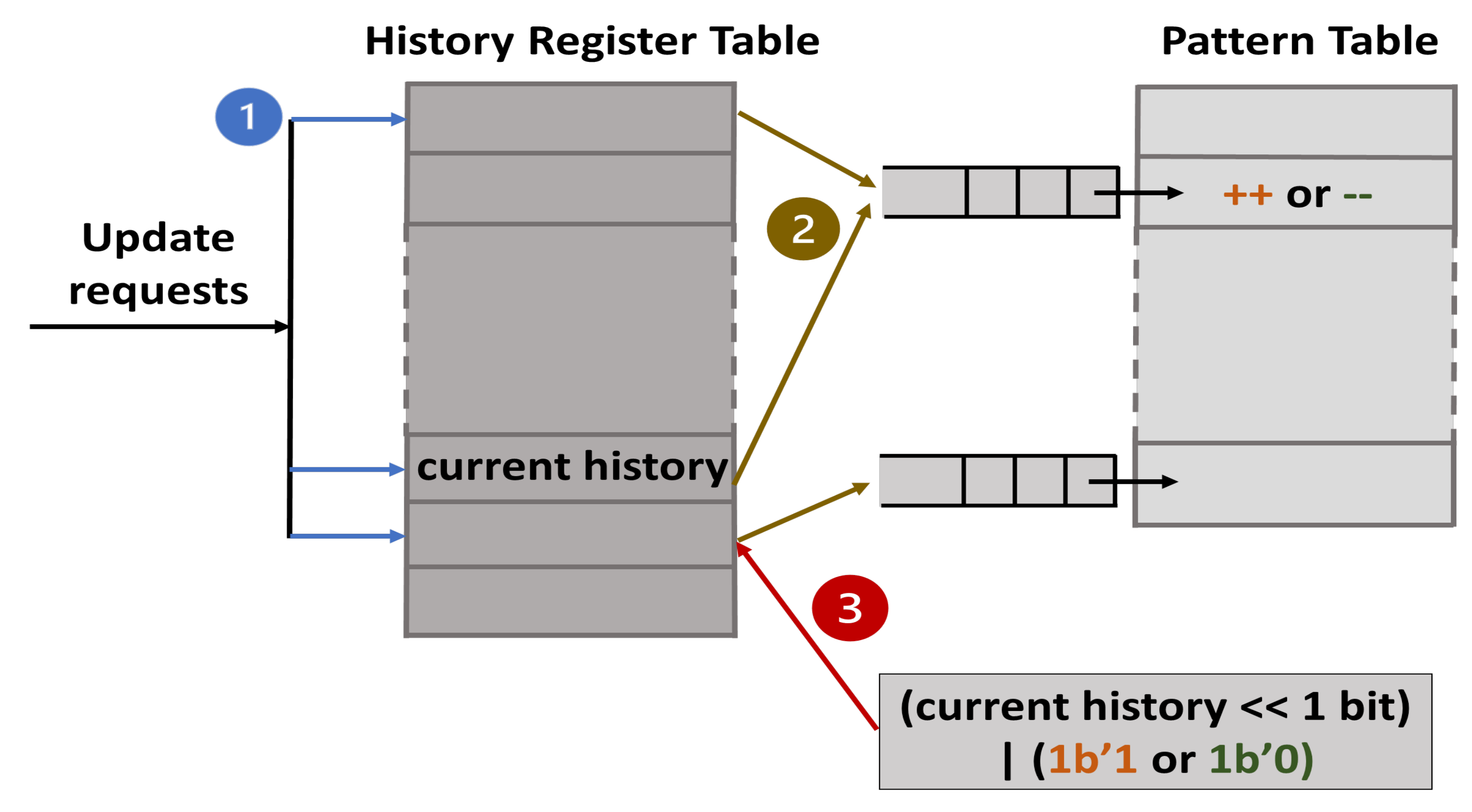}}
  \caption{Updating predictor tables}
  \label{update_predictor}    
\end{center}
\end{figure}

When an instruction block is being searched for in a CSHR set, it can match the i-Filter block field of at most one CSHR entry. This is because when a block becomes i-Filter victim again and is inserted into the CSHR set, it must have already been re-accessed for it to have got back into the i-Filter after the previous eviction. This re-access guarantees that the block's previous comparison has been resolved and the corresponding CSHR entry is no longer valid, if we assume updating the predictor tables can finish before the block becomes i-Filter victim again (which does occur in most cases, as described in the next paragraph). However, the instruction block being searched can match the i-cache contender block field of multiple CSHR entries, because the i-cache contender block can be compared with different i-Filter victims and wins the competition each time to stay in i-cache. Therefore, one instruction block being searched can lead to multiple HRT and PT update requests. To address this, HRT is first indexed in parallel, and then the current history values in HRT are used to index and update PT in the next cycle. 
After the history values are passed to the PT updater, the current history registers in HRT are updated accordingly. Aliasing can occur when updating HRT and PT, and can cause conflicts when we update multiple entries in parallel. However, we find that aliasing in indexing HRT is so rare that we simply update each HRT entry for only one request and ignore the others. Since PT is much smaller than HRT, the probability of aliasing in PT is a little higher. To mitigate this problem, we add a 10-slot queue for each PT entry to accommodate the update requests. In each cycle, the heads of the PT update queues are popped and are used to update the PT entries. Figure~\ref{update_predictor} summarizes the datapath to update the predictor tables after matched CSHR entries are found.

\begin{figure}[h!]
\begin{center}
  \includegraphics[width=\linewidth]{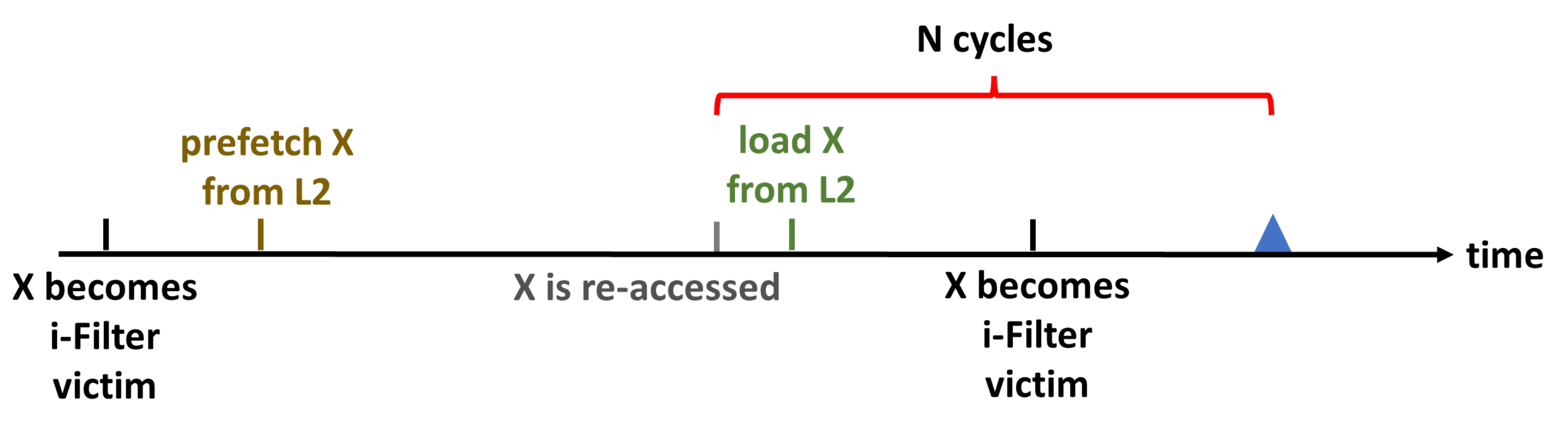}
  \caption{CSHR entry could not be resolved in time with a  prefetcher}
  \vspace{-0.5cm}
  \label{timeline_CSHR}    
\end{center}
\end{figure}

There is a concern that due to the 2 cycles (or more if waiting in the PT entry update queue) spent in updating the predictor tables, a block \emph{X} may become i-Filter victim again while there is already one unresolved CSHR entry whose i-Filter victim block field is \emph{X}.
This implies that the stale (older) information for block \emph{X} in the predictor tables is used to make prediction this time. We illustrate this problem in Figure~\ref{timeline_CSHR}. 

In the case with a prefetcher, where block \emph{X} is prefetched before it is re-accessed, the predictor could be updated after block \emph{X} becomes i-Filter victim again. As shown in the timeline, the prefetch request reduces the cycles between re-accessing block \emph{X} and loading it into i-Filter from L2 cache. For a superscalar processor, it could take as few as 3 cycles for block \emph{X} to move from the MRU position to LRU position in i-Filter. Therefore, it is possible that the CSHR entry could not be resolved in time with the presence of a prefetcher if \emph{$N > 3$}. However, as discussed in Section~\ref{ssec:update_latency}, such cases have a negligible impact on the overall performance.

\vspace{-0.2cm}
\subsection{Additional Storage and Energy requirements for ACIC}\label{ssec:storage_requirement}
\vspace{-0.1cm}
\noindent {\em Storage:}
Each i-Filter entry contains 58 tag bits, 1 valid bit, 4 LRU bits, which adds up to 63 metadata bits, and a 64B instruction block. We empirally determine the size of HRT to be 1024 entries, each of which consists of 4 history bits, leading to $2^4$ entries in PT. Each PT entry contains a 5-bit counter that indicates the prediction result. Each of the 10 slots in PT entry update queue contains a 4-bit PT index and 1 bit indicating whether the counter in the PT entry should be incremented or decremented.

CSHR contains 256 entries, and each entry consists of 12 tag bits for the i-Filter victim block, 12 tag bits for the i-cache contender block, 1 valid bit, and 5 LRU bits for the 32-way CSHR design. Table~\ref{table:storage_overhead} summarizes the storage overhead of ACIC for a 32KB i-cache with 8-way associativity. Evaluation results of ACIC in Section~\ref{evaluation} are based on the ACIC parameters in Table~\ref{table:storage_overhead}, and we provide sensitivity analysis of ACIC in Section~\ref{ssec:sensitivity}.

We also list the storage overhead of the prior schemes that we compared with in Table~\ref{table:prior_works_summary}. ACIC requires 2.67KB extra storage, which is roughly 2/3rd of the 4.06KB storage overhead of GHRP, the state-of-the-art i-cache replacement policy with hardware techniques. 

\begin{scriptsize}
\begin{table}[h!]
  \centering
  \caption{Storage overhead of ACIC for a 32KB, 8-way i-cache}
  \label{table:storage_overhead}
  \begin{tabular}{|p{0.30\linewidth} | p{0.60\linewidth}|}
    \hline
    \textbf{Component} & \textbf{Number of bits} \\
    \hline
    \hline
    i-Filter & 16 entries $\times$ (63 bit metadata + 64B instruction block) = 1.123KB \\
    \hline
    HRT & 1024 entries $\times$ 4 bit history = 0.5KB \\
    \hline
    PT & $2^4$ entries $\times$ 5 bit counters = 10B\\
    \hline
    PT entry update queue & 16 PT update queues $\times$ 10 slots $\times$ (4 bit PT idx + 1 bit update request) = 100B \\
    \hline
    CSHR & 256 entries $\times$ (24 bit tags + 1 bit valid + 5 bit LRU) = 0.9375KB \\
    \hline
    \textbf{Total} & \textbf{2.67KB}  \\
    \hline
  \end{tabular}
\end{table}
\end{scriptsize}

\noindent {\em Energy:}
We use the power pack (of the simulation infrastructure described in Section~\ref{simulation-infrastructure}) to measure the chip energy for a 22nm process technology. It uses the McPAT~\cite{McPAT} model, and we calculate power for the i-Filter, HRT, PT, and CSHR with CACTI 7~\cite{cacti-7} and add the estimated values to the McPAT power numbers. It includes the chip energy with total execution time, runtime dynamic power, and total leakage power. We find that {\em ACIC saves 0.63\% chip energy on average}, despite the additional power taken by the new structures. While this is only the chip energy, the higher speedup and higher i-cache hit rates of ACIC, will further decrease the overall system energy if we consider off-chip DRAM, interconnects and peripherals.

\section{Evaluation}\label{evaluation}
\subsection{Simulation infrastructure}\label{simulation-infrastructure}
\begin{scriptsize}
\begin{table}[h!]
  \centering
  \caption{Simulation parameters}
  \label{table:parameters}
  \begin{tabular}{p{0.30\linewidth}  p{0.60\linewidth}}
    \hline
    \textbf{Parameter} & \textbf{Value}\\
    \hline
    CPU frequency & 4GHz \\
    Fetch width & 6-wide, 24-entry Fetch Target Queue \\
    Decode width & 6-wide, 60-entry Decode Queue \\
    Out-of-order Core & 352-entry Reorder Buffer \\
    \hline
    BTB & 8192-entry, 4-way \\
    Branch predictor & TAGE~\cite{TAGE} \\
    \hline
    L1 I-Cache & 32KB, 8-way, 16 MSHRs, 4-cycle \\
    L1 D-Cache & 48KB, 8-way, 16 MSHRs, 5-cycle \\
    L2 Unified Cache & 512KB, 8-way, 32 MSHRs, 15-cycle \\
    L3 Unified Cache & 2MB, 16-way, 64 MSHRs, 35-cycle \\
    DRAM & 1 channel, 3200MT/s (25.6GB/s) \\
    \hline
  \end{tabular}
\end{table}
\end{scriptsize}

We first collect the full system execution trace of each application with the Qemu~\cite{qemu} emulator. Specifically, a trace of 500 million or 1 billion instructions (depending on the execution time of the application) in the steady state is recorded. The traces are then fed to the Tejas~\cite{tejas} simulator, a detailed cycle accurate trace-driven simulator. In each simulation, the simulator is warmed up with the first 10\% (i.e. 50-100 million) of the instructions. Our core model is similar to the Intel Sunny Cove, as shown in Table~\ref{table:parameters}. 

\begin{scriptsize}
\begin{table}[h]
  \centering
  \caption{Data center applications used in our evaluation}
  \label{table:benchmarks}
  \begin{tabular}{|p{0.18\linewidth}|p{0.20\linewidth} |p{0.32\linewidth}|p{0.12\linewidth}|}
    \hline
    & \textbf{Benchmark Suite} & \textbf{Description} & \textbf{MPKI} \\
    \hline
    \hline
    Media Streaming & CloudSuite\newline~\cite{cloudsuite} & Darwin streaming server & 81.2  \\
    \hline
    Data Caching & CloudSuite\newline~\cite{cloudsuite} & Memcached for Twitter & 78.1 \\
    \hline
    Data Serving & CloudSuite\newline~\cite{cloudsuite} & YCSB data store server & 31.6 \\
    \hline
    Web Serving & CloudSuite\newline~\cite{cloudsuite} & cloud web services & 65.8 \\
    \hline
    Web Search & CloudSuite\newline~\cite{cloudsuite} & Apache Solr search engine & 151.5 \\
    \hline
    TPC-C & OLTP-Bench~\cite{oltpbench} & OLTP workload & 42.5 \\
    \hline
    Wikipedia & OLTP-Bench~\cite{oltpbench} & online encyclopedia & 41.1 \\
    \hline
    SIBench & OLTP-Bench~\cite{oltpbench} & snapshot isolations in DBMSs & 35.0 \\
    \hline
    Finagle-HTTP & Renaissance\newline~\cite{renaissance} & Twitter's HTTP server & 46.1 \\
    \hline
    Neo4J-Analytics & Renaissance\newline~\cite{renaissance} & graph queries for a database & 58.7 \\
    \hline
  \end{tabular}
\end{table}
\end{scriptsize}

\subsection{Prior Works for Comparison}
ACIC has similar motivations (avoiding and dealing with i-cache pollution) targeted by the following three broad strategies: cache replacement policies, cache bypassing policies, and victim cache. Consequently, we compare ACIC quantitatively with prior and recent proposals that fall in these three categories as shown in  Table~\ref{table:prior_works_summary}. The \textit{Cache Type} column identifies the cache targetted by the original proposal. For each of these prior proposals, we also list their important parameters used in the simulations, along with the additional storage that they require. As Table~\ref{table:prior_works_summary} shows, for the simulated system, ACIC imposes an additional storage requirement of 2.67KB, which is around 2/3rd of the recent GHRP~\cite{GHRP} proposal.

Additionally, a prefetcher, which reduces i-cache misses, can complement or belittle the benefits of these prior/our proposals. Consequently, we consider a standard fetch-directed prefetcher (FDP)~\cite{FDP}.

\begin{scriptsize}
\begin{table*}[ht]
  \centering
  \caption{Schemes for comparison}
  \label{table:prior_works_summary}
  \begin{tabular}{|p{0.12\linewidth}|p{0.15\linewidth}|p{0.09\linewidth}|p{0.45\linewidth}|p{0.09\linewidth}|}
    \hline
     & \textbf{Optimization Strategy} & \textbf{Cache Type} & \textbf{Important Parameters/Notes} & \textbf{Storage Overhead}\\
    \hline
    \hline
    SRRIP~\cite{SRRIP} & replacement policy & LLC & 2-bit RRPV & 0.125KB \\
    \hline
    SHiP~\cite{SHiP} & replacement policy & LLC & 13-bit signature, 8K-entry SHCT, 2-bit counter & 2.88KB\\
    \hline
    Hawkeye~\cite{Hawkeye}/\newline Harmony~\cite{Harmony} & replacement policy & LLC & 64 entries per occupancy vector, 8K-entry predictor, 3-bit training counter, 3-bit RRIP & 4.69KB\\
    \hline
    GHRP~\cite{GHRP} & replacement policy & L1 i-cache & 3 4096-entry predictor tables, 2-bit counter, 16-bit signature, 1-bit prediction, 16-bit history register & 4.06KB\\
    \hline
    DSB~\cite{DSB} & bypassing policy & LLC & 16-bit tracked line tag, 3-bit competitor way tag, 2 sampled sets for policy selection & 0.48KB \\
    \hline
    OBM~\cite{OBM} & bypassing policy & LLC & 21-bit incoming block tag, 21-bit victim block tag, 10-bit signature, 128-entry RHT, 1024-entry BDCT, 4-bit counter & 1.41KB \\
    \hline
    VVC~\cite{VVC} & victim cache & LLC & 15-bit trace, 2 $2^{14}$-entry predictor tables, 2-bit counter & 9.06KB \\
    \hline
    VC8K~\cite{victim-cache} & victim cache & L1 cache & 4-way associative, 128 blocks & 8KB \\
    \hline
    40KB i-cache & larger i-cache & L1 i-cache & 10-way associative, 640 blocks & 8KB \\
    \hline
    OPT~\cite{belady} & replacement policy & all types & evict the block that is reused furthest in the future & 0KB \\
    \hline
    OPT bypass with i-Filter & bypassing policy & L1 i-cache & place i-Filter victim in i-cache only if i-Filter victim is known (with oracle knowledge) to have smaller reuse distance than the i-cache contender selected by LRU  & 1.123KB \\
    \hline 
    ACIC & bypassing policy & L1 i-cache & 16-entry i-Filter, 1024-entry HRT, 4-bit history, $2^4$-entry PT, 5-bit counter, 10-entry PT entry update queue, 256-entry CSHR, 24-bit partial tags & 2.67KB \\
    \hline
  \end{tabular}
\end{table*}
\end{scriptsize}

\subsection{Workloads and Metrics}
Table~\ref{table:benchmarks} lists the datacenter applications used in our evaluations. These applications have been noted to suffer from front-end bottlenecks in related studies~\cite{Shotgun, divide-and-conquer, Ripple} due to their large footprints, involvement of libraries and OS, as well as varying dynamism in their execution paths. 
Column \emph{MPKI} quantifies the i-cache MPKI (misses per 1000 instructions) in these applications on our FDP baseline platform.

The most important metric for an application is the execution time, and speedup of any proposed enhancement over the baseline is the first metric that we consider. Equally important is the reduction in i-cache misses (MPKI) attained with the enhancements, since those are the key targets of optimization in these schemes. Consequently, we study both these metrics in our evaluation below.

\subsection{Comparison with replacement policies (SRRIP, SHiP, Hawkeye/Harmony, GHRP)}\label{ssec:replacement_policies_comparison}
From Figure~\ref{speedup_comparisons}, we can see that the recently proposed GHRP provides the highest speedup amongst these previously proposed replacement policies. Still, ACIC outperforms GHRP with FDP. In particular, ACIC provides 1.0223 speedup on average over the LRU replacement policy FDP baseline, which corresponds to 56.03\% of the attainable speedups of the oracle-based OPT replacement policy. 

GHRP uses instruction reuse to predict dead blocks in the i-cache and prioritizes such dead blocks for replacement. If we define replacement accuracy as the percentage of victims selected by a given policy (e.g. GHRP) that are identical to the victims selected by OPT, we find that the replacement accuracy of GHRP is 17.90\% on average, resulting in 15.64\% of the MPKI reduction provided by OPT. ACIC is much more accurate, reducing 55.85\% of misses reduced by OPT, as shown in Figure~\ref{MPKI_reduction_comparisons}.

As can be seen from Figure~\ref{MPKI_reduction_comparisons}, \emph{Media streaming}, \emph{Data caching}, \emph{Web search}, and \emph{Neo4J-analytics} are applications that show higher MPKI reduction under ACIC and GHRP than the other applications. The potential of a replacement policy is determined by the performance/MPKI difference between the OPT replacement policy and the baseline LRU policy. With the larger headroom, these four applications, ACIC and GHRP can help them to a greater extent. These are also those applications which suffer more from the burstiness behavior identified earlier (Figure~\ref{reuse_dist_distribution}), for which LRU cannot predict and optimize for the larger reuse distance after a recent burst. The relative benefits across applications with ACIC is further explained in Section~\ref{ssec:ACIC_insights}.

\subsection{Comparison with bypassing policies (DSB and OBM)}
Of these two prior bypassing policies, DSB  performs slightly better, though providing only a limited 1.0006 speedup over the LRU baseline with FDP.

DSB bypasses newly allocated blocks from the cache with a probability tuned based on the effectiveness of past bypassing decisions. 
Though similar in goals, unlike ACIC, DSB does not provide spatio-temporal locality separation whose importance was pointed out in Section~\ref{motivation}.
Even with a higher storage budget, DSB does not perform as well as ACIC due to this fundamental problem. DSB provides only 0.46\% MPKI reduction over the LRU baseline on average. Moreover, the bypassing policy of DSB is not very effective, and when equipped with i-Filter, DSB still only provides 1.0010 speedup over baseline.

Interestingly, we see that the results for OPT bypassing and OPT replacement are similar/close, implying that combining the spatio-temporal separation provided by i-Filter and a good admission control mechanism, can be an effective way to improve i-cache performance.
\color{black}

\subsection{Comparison with victim caches (VC3K, VVC) and larger i-cache}\label{36KB_and_VC3K}
\vspace{-0.1cm}
One could question whether the real-estate required for the filtering mechanism could have been better served with an appropriate victim cache (which temporarily retains evictions for another chance), or even a larger i-cache.
Consequently, we compare ACIC with (i) a traditional 3KB fully-associative victim cache VC3K~\cite{victim-cache}, (ii) a recent work on victim cache~\cite{VVC}, VVC which better uses the existing space,  and (iii) a larger 36KB, 9-way i-cache (i.e. adding 4KB over our baseline, which is more than the additional real-estate needed for ACIC and also has a higher associativity). 

\begin{figure*}[!t]
\centering
  \includegraphics[width=\linewidth]{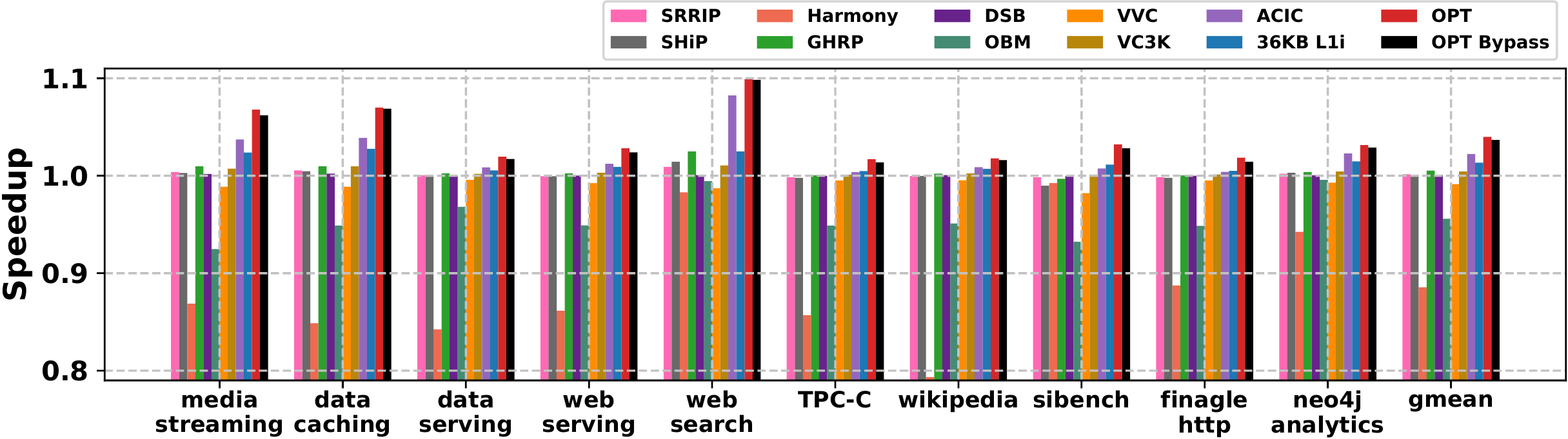}
  \caption{ACIC's speedup compared with state-of-the-art replacement, bypassing, and victim cache policies over an LRU baseline with fetch-directed prefetching.}
  \label{speedup_comparisons} 
\end{figure*}

VVC turns out to actually slow down the execution as seen in Figure~\ref{speedup_comparisons}. VVC uses slots in the existing i-cache that are predicted dead to hold blocks evicted from other sets. We find that in nearly 60\% of the cases, the victim blocks have longer reuse distances than the predicted dead blocks in other sets, but they are still brought into other sets by VVC, leading to waste of cache capacity. 
While a traditional victim cache (VC3K) does much better than VVC, ACIC
gives 1.018$\times$ the speedup provided by the 3KB victim cache on average. 

Figure~\ref{speedup_comparisons} shows that ACIC provides  1.009$\times$ the speedup provided by the 36KB i-cache on average. These results tend to reiterate the importance of being more discretionary in what comes into and goes out of i-cache, than blindly throwing more resources at it. 

\begin{figure*}[!t]
\centering
  \includegraphics[width=\linewidth]{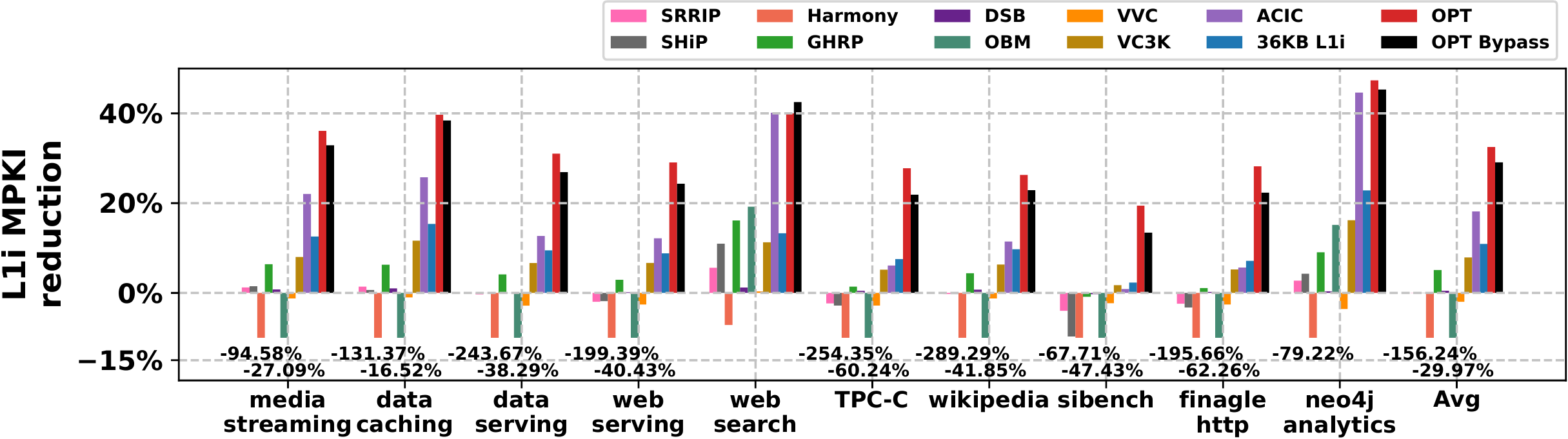}
  \caption{ACIC's MPKI reduction compared with state-of-the-art replacement, bypassing, and victim cache policies over an LRU baseline with fetch-directed prefetching.}
  \label{MPKI_reduction_comparisons} 
\end{figure*}

\subsection{Insights into the working of ACIC}\label{ssec:ACIC_insights}
\begin{figure*}[!t]
    \centering
  \subfloat[\normalsize Average ACIC bypass accuracy for various reuse distance ranges
 \label{ACIC_bypass_accuracy}]{
  \includegraphics[width=0.48\linewidth ]{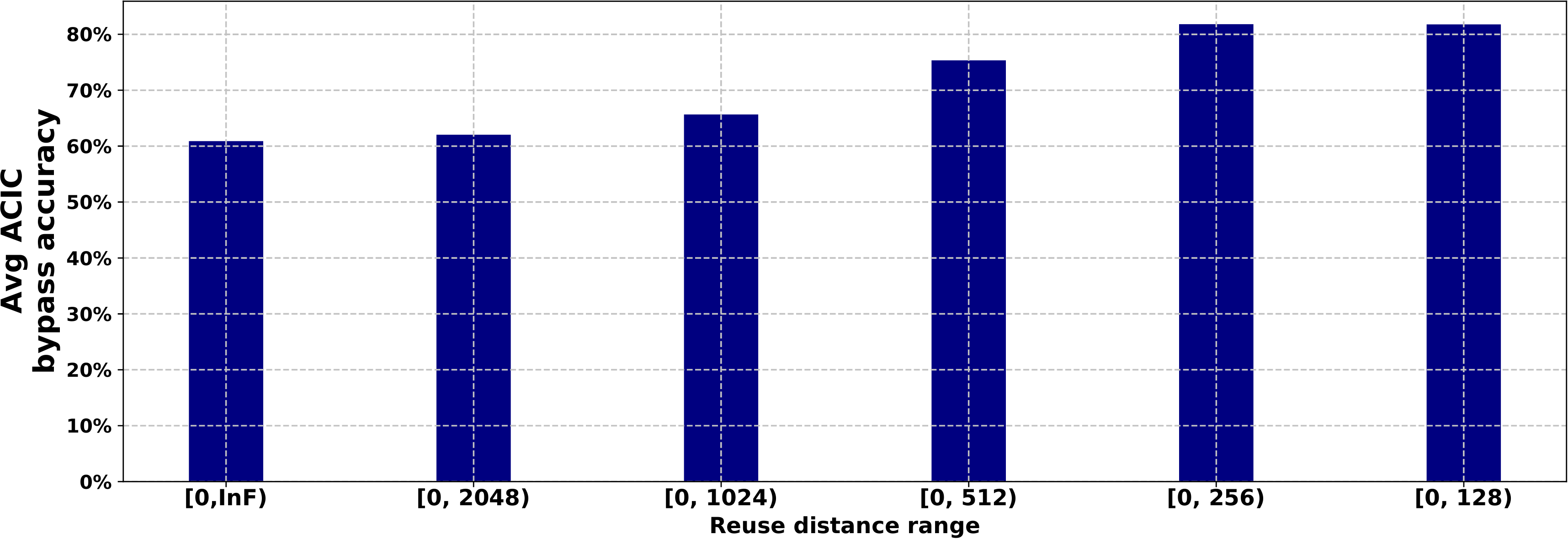}}
  \hspace{0.2cm}
 \subfloat[\normalsize MPKI reduction comparison of random bypass with 60\% accuracy and ACIC over fetch-directed prefetching baseline
\label{random_bypass_vs_ACIC}]{
  \includegraphics[width=0.485\linewidth]{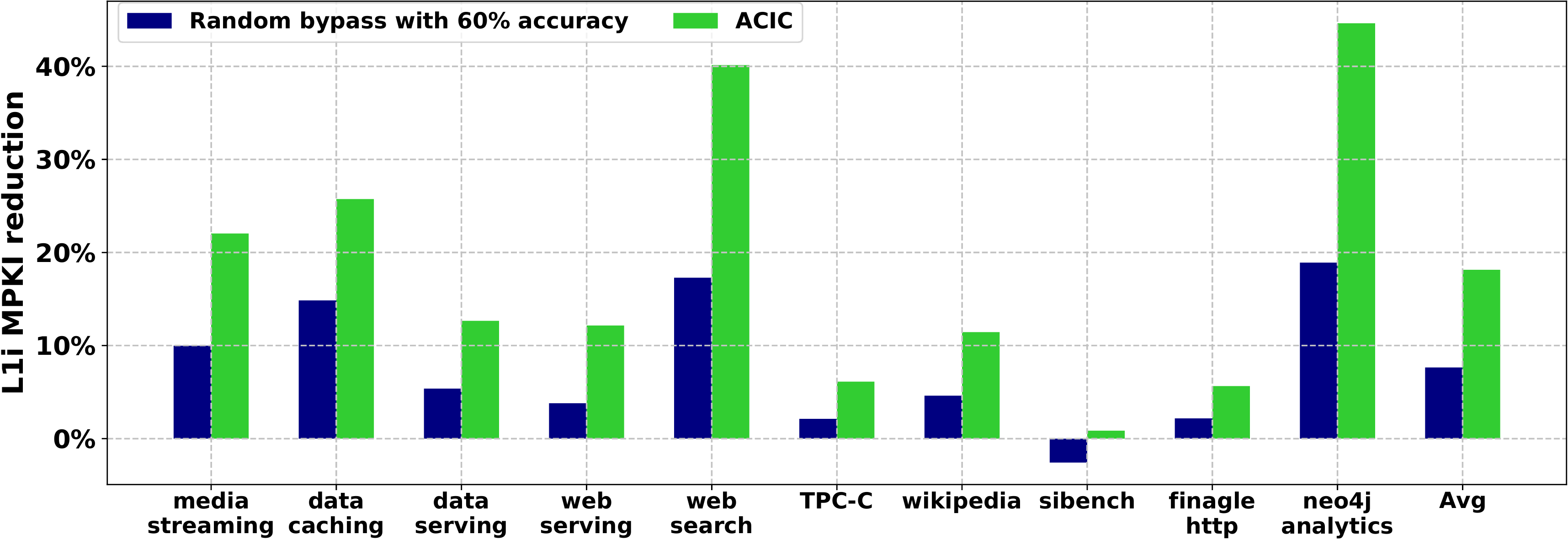}}
  \caption{ACIC bypass accuracy analysis}
  \label{ACIC_accuracy} 
\end{figure*}

\begin{figure}[h!]
\begin{center}
  \includegraphics[width=\linewidth]{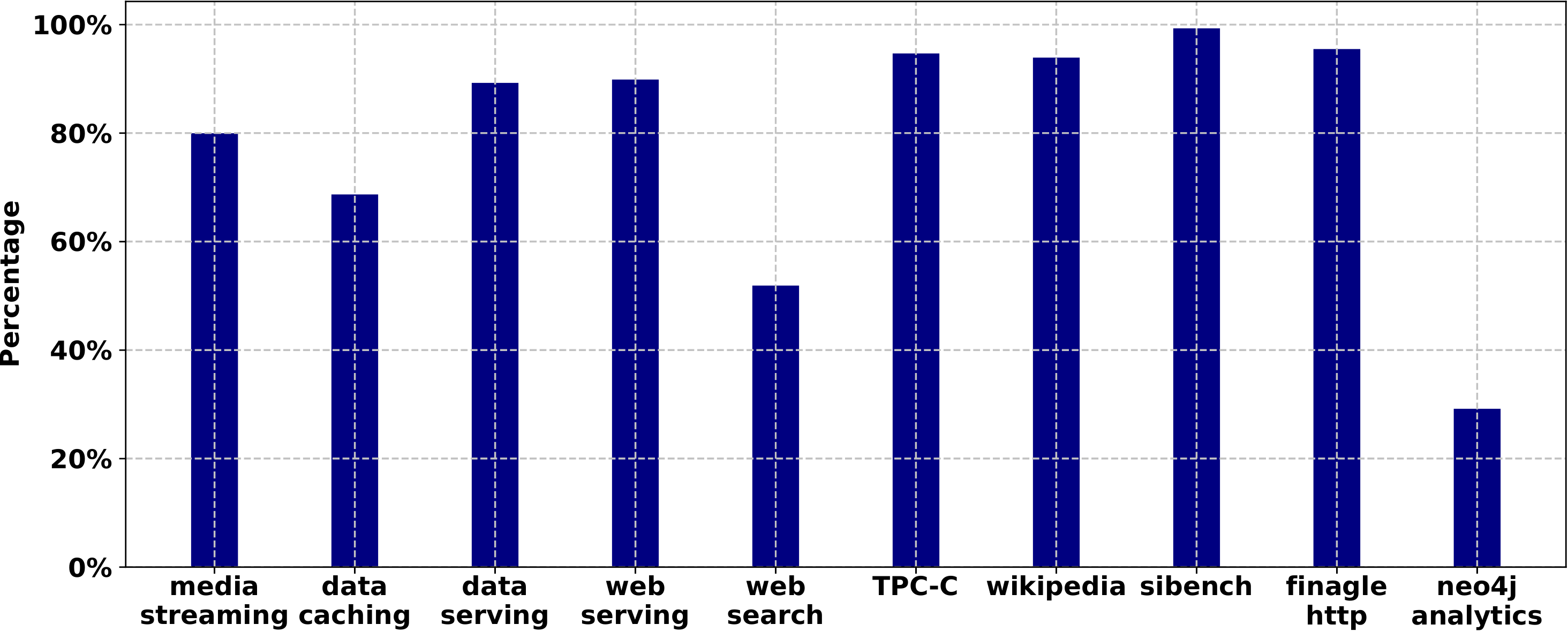}
  \caption{Percentage of i-Filter victims inserted into i-cache}
  \label{evict_to_L1i_percentage}    
\end{center}
\vspace{-0.4cm}
\end{figure}

\noindent {\em Discretionary Filtering:} Figure~\ref{evict_to_L1i_percentage} depicts the percentage of i-Filter victims that are inserted into i-cache based on the predictor in ACIC. The percentages vary significantly across applications (from 30-99\%). As Figure~\ref{reuse_dist_distribution} showed, \emph{Web search}, \emph{Neo4J-analytics}, \emph{Data caching}, and \emph{Media streaming} show a higher fraction of reuse distances which fall just beyond the i-cache's reach, where it becomes more critical to decide whether or not to insert the victim from i-Filter into i-cache. This is confirmed by Figure~\ref{evict_to_L1i_percentage}, where we see these applications exhibiting a larger filtering effect. This reiterates the need for dynamic adaptation to application behavior as in ACIC, rather than a static way of determining whether to insert into i-cache after the current burst. 

\noindent {\em Accuracy of Filtering:}
It is even more important to examine whether ACIC made the correct filtering choice. To do this, we use oracle knowledge about reuse distances to compare the future reuse distances of the i-cache victim and the i-Filter victim, and compare that decision with ACIC's prediction. The filter accuracy of ACIC is calculated as the percentage of the correct predictions over total predictions. Surprisingly,
the average bypass accuracy of ACIC is only 60.89\%, as shown in the first bar (corresponding to [0,InF)) of Figure~\ref{ACIC_bypass_accuracy}. However, the bypass accuracy matters only in cases when the reuse distances of the i-Filter victim and the i-cache contender block are not both very large (if they are, they will both likely get evicted before being accessed), and their reuse distances are not equal either. We consequently plot the ACIC bypass accuracy for varying ranges of reuse distances in Figure~\ref{ACIC_bypass_accuracy}.

To demonstrate that ACIC is reasonably accurate where it really matters, we also consider a ``random'' filtering mechanism to determine whether to insert the evicted i-Filter block into i-cache. In  Figure~\ref{random_bypass_vs_ACIC}, we compare the i-cache MPKI reduction of ACIC and this random bypass scheme over the FDP baseline. Even though the random bypass scheme has 60\% accuracy, similar to the overall bypass accuracy of ACIC, we can see that it provides only 7.65\% MPKI reduction, which is 42.17\% of the MPKI reduction provided by ACIC. Figure~\ref{ACIC_bypass_accuracy} and Figure~\ref{random_bypass_vs_ACIC} provide a key insight: prediction accuracy matters only when at least either of the two (i-Filter victim or i-cache contender) has a reuse distance that is not very large, so that at least one of them is likely to be accessed again while in i-cache in the near future.

\noindent {\em Latency in updating predictor:}\label{ssec:update_latency}
\begin{figure}[h!]
\begin{center}
  \includegraphics[width=\linewidth]{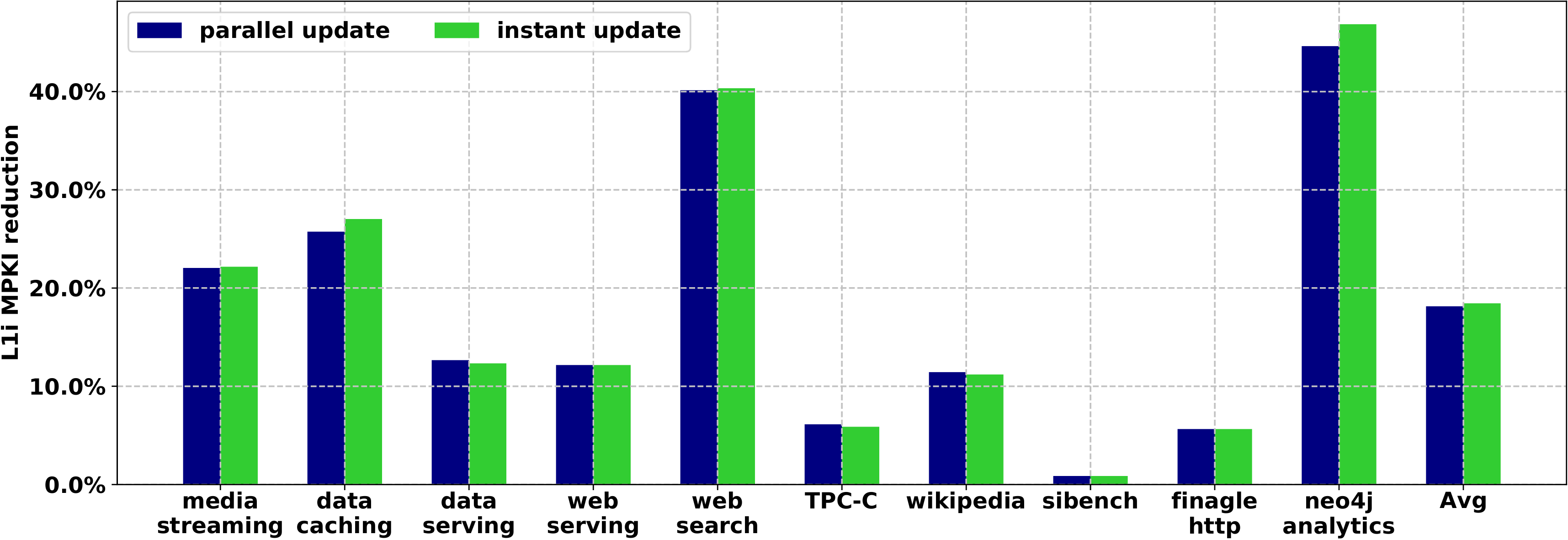}
  \caption{MPKI reduction for instantly updating vs. 2-cycle latency (parallel) in updating CSHR}
  \label{parallel_vs_instant_update}    
\end{center}
\end{figure}
Section~\ref{sssec:access_cycles} described the possibility that stale information is read from predictor due to the multiple cycles spent in updating the two tables, HRT and PT, with the existence of a prefetcher. To see whether this could cause a problem in performance, we compare the i-cache MPKI reduction with our parallel update scheme, in which at least 2 cycles are spent in updating HRT and PT, and an instant update scheme, in which the HRT and PT are updated immediately. From Figure~\ref{parallel_vs_instant_update}, we can see that the MPKI reduction of the parallel update scheme is very close to that of the instant update scheme. The update latency of the predictor tables thus does not affect ACIC's effectiveness, and does not need to come into the critical path.

\noindent {\em Sensitivity Analysis:}\label{ssec:sensitivity}
\begin{figure}[h!]
\begin{center}
  \includegraphics[width=\linewidth]{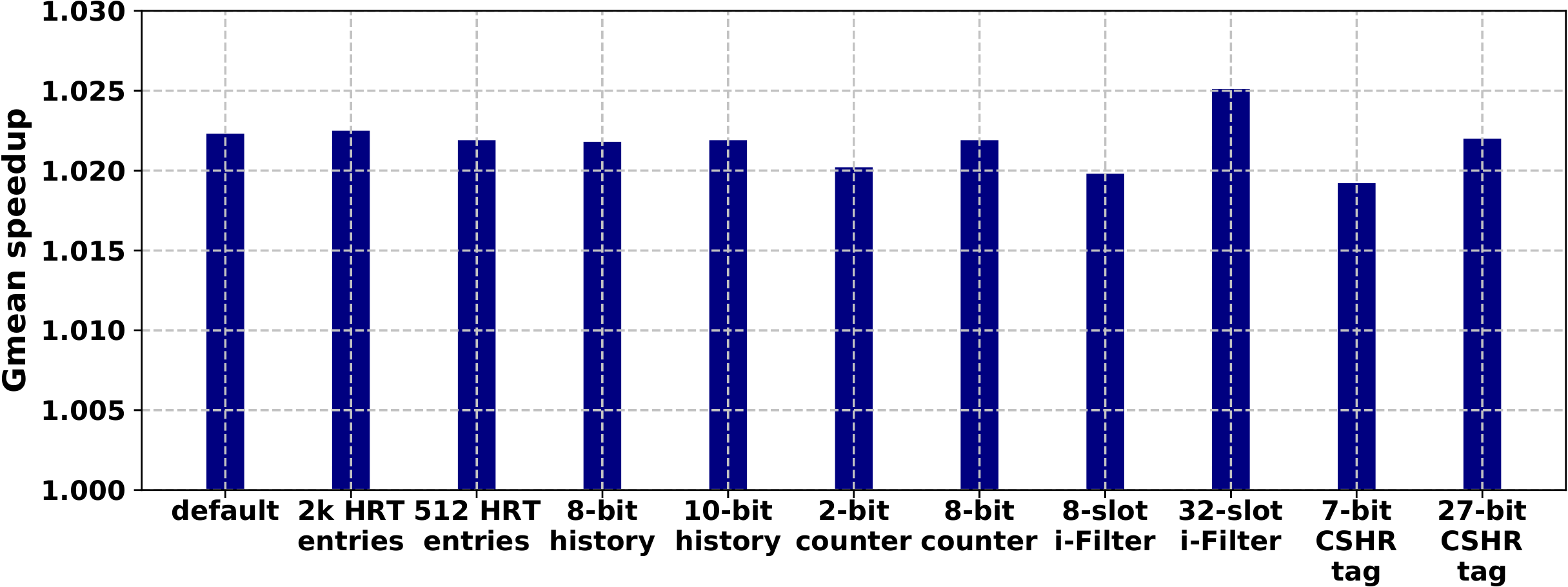}
  \vspace{-0.5cm}
  \caption{Sensitivity of ACIC to different configurations}
  \vspace{-0.3cm}
  \label{ACIC_sensitivity}    
\end{center}
\end{figure}
Figure~\ref{ACIC_sensitivity} shows the average speedup of ACIC when its key design parameters are varied. The leftmost bar \emph{default} gives the average speedup of ACIC with parameters shown in Table~\ref{table:storage_overhead}. Since the number of CSHR entries has been discussed in Section~\ref{sssec:CSHR_size}, here we only show sensitivity to HRT entries, length of each history register in HRT, length of counters in PT, number of i-Filter slots, and length of partial tags in CSHR. We can see that among all the parameters, increasing the i-Filter size gives the most benefit, while decreasing i-Filter size, length of PT counter and CSHR tags worsen performance the most. Increasing the history length from 4-bit to 10-bit does not show a big performance gain. 

\subsection{Discussion}
\subsubsection{Performance benefit due to bypass policy}\label{ACIC_over_ibuf_baseline}
\begin{figure}[h!]
\begin{center}
  \includegraphics[width=\linewidth]{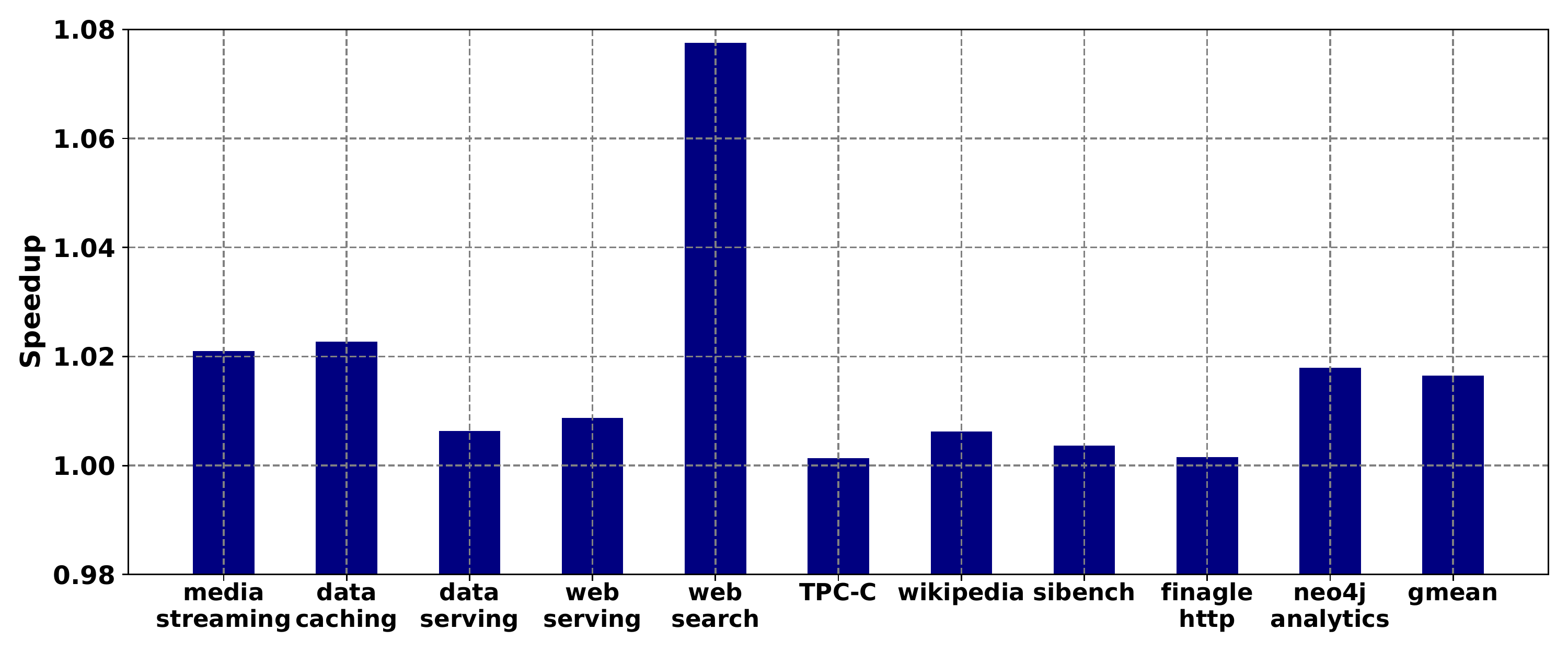}
  \caption{ACIC speedup over FDP baseline with i-Filter}
  \label{ACIC_speedup_over_ibuf_baseline}    
\end{center}
\end{figure}
While similar/variant forms of i-Filter are not necessarily modeled in current academic simulators, i-Filter-like small buffers are usually present in real processors to contain recently accessed instruction blocks. To show the benefit of ACIC more realistically, we present Figure~\ref{ACIC_speedup_over_ibuf_baseline} to show the speedup of ACIC over FDP baseline equipped with i-Filter. We can see that ACIC's bypass policy itself gives 1.0165 geomean speedup over the LRU replacement policy baseline.

\subsubsection{Necessity of each ACIC structure}\label{necessity_of_ACIC_parts}
\begin{figure}[h!]
\begin{center}
  \includegraphics[width=\linewidth]{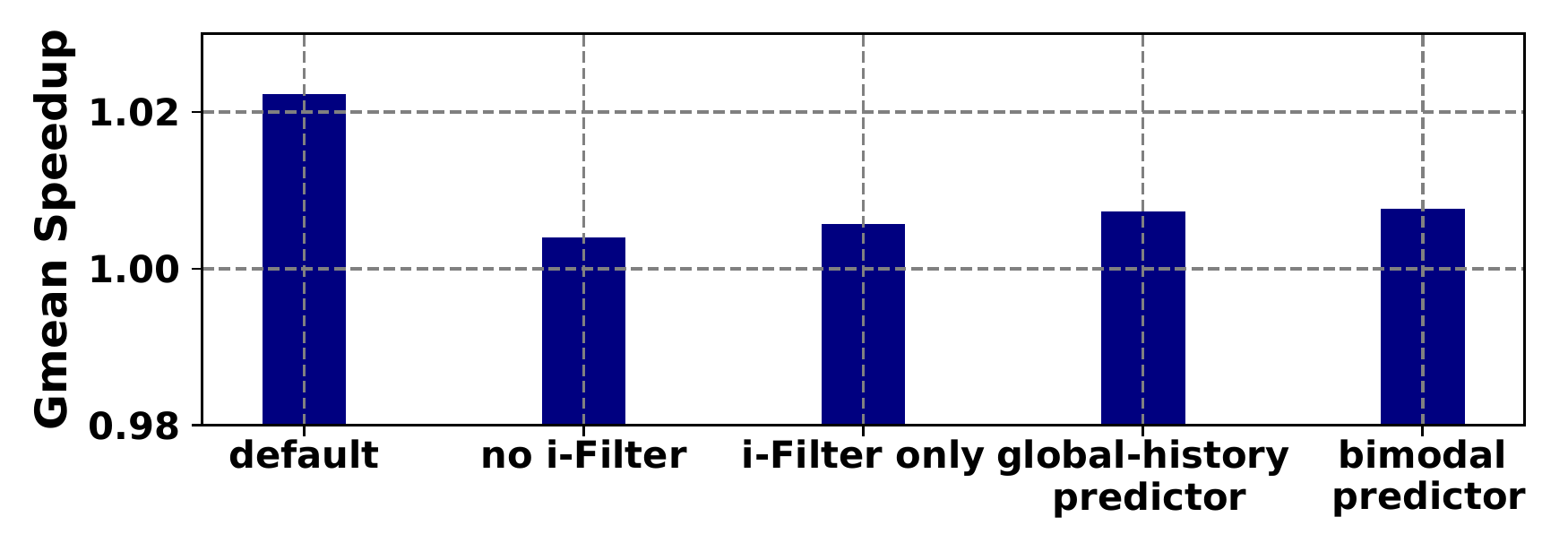}
  \caption{Speedup of ACIC with simpler designs over FDP baseline}
  \label{turn_off_ACIC_parts_speedup}    
\end{center}
\end{figure}
While ACIC gives better performance and less storage overhead than the recently proposed GHRP, ACIC's mechanism is more complex. CSHR is responsible for training the predictor, so it cannot exist on its own. To justify the necessity of the other two parts of ACIC (i-Filter and two-level predictor), we plot Figure~\ref{turn_off_ACIC_parts_speedup} to show the geomean speedup of ACIC with simpler designs over FDP baseline: ACIC without i-Filter, ACIC with i-Filter only, ACIC with a global history two-level predictor, and ACIC with a bimodal predictor. We can see that turning off i-Filter/predictor or replacing two-level predictor with simpler ones does not give as good performance as our default ACIC.

\subsubsection{Evaluation of ACIC with SPEC workloads}\label{spec_evaluation}
\begin{figure}[h!]
\begin{center}
  \includegraphics[width=\linewidth]{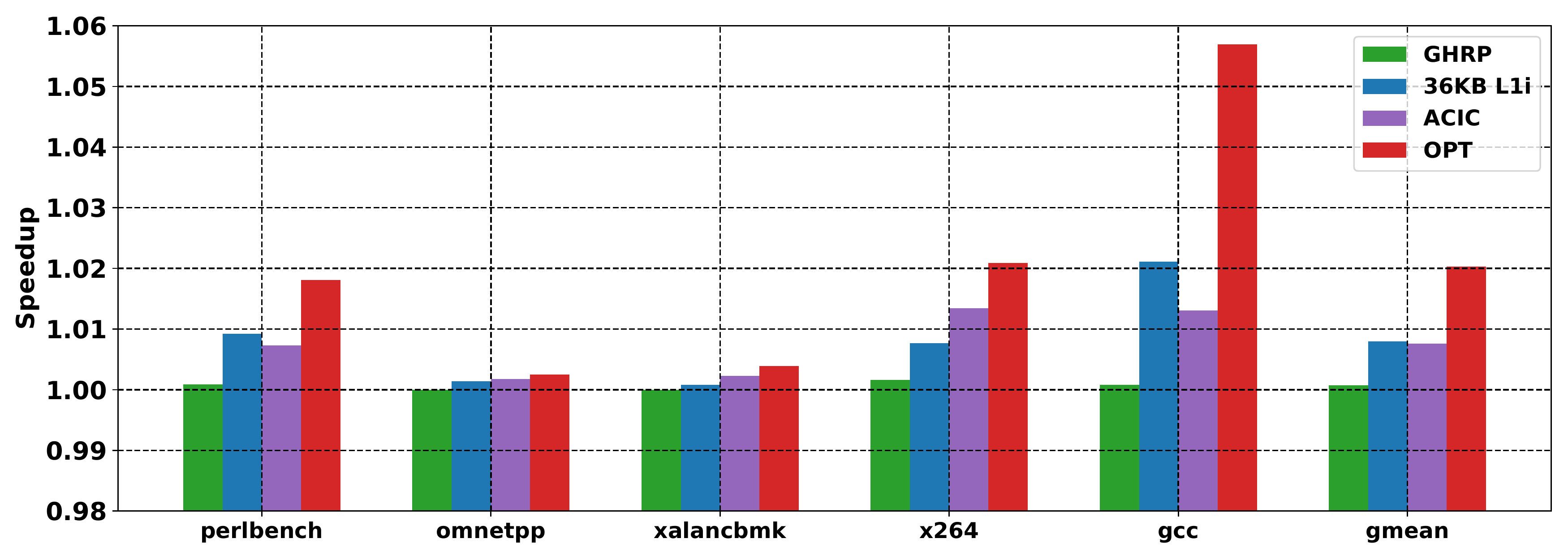}
  \caption{Speedup comparison of various policies over FDP baseline for SPEC workloads}
  \label{spec_speedup_comparisons}    
\end{center}
\end{figure}

\begin{figure}[h!]
\begin{center}
  \includegraphics[width=\linewidth]{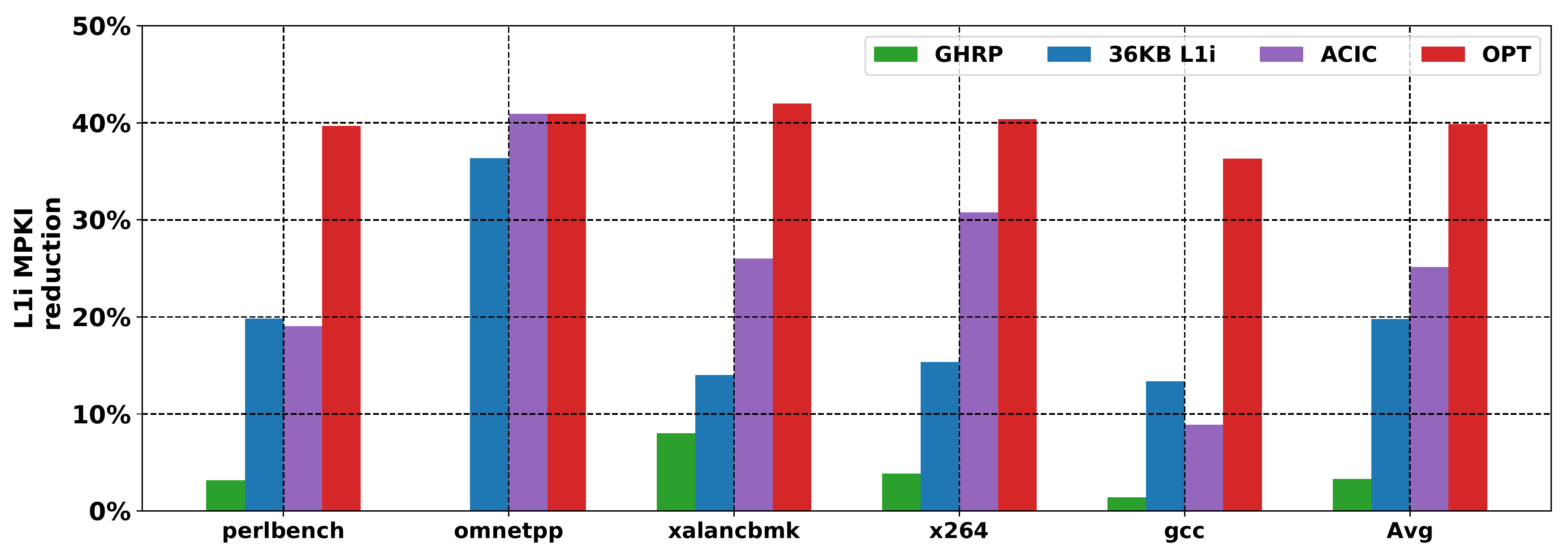}
  \caption{MPKI reduction comparison of various policies over FDP baseline for SPEC workloads}
  \label{spec_MPKI_reduction_comparisons}    
\end{center}
\end{figure}
ACIC targets datacenter workloads, as these workloads suffer from higher i-cache misses than conventional workloads like SPEC~\cite{spec2017}. For completeness, we evaluate how ACIC performs in SPEC workloads as well. Figure~\ref{spec_speedup_comparisons} and Figure~\ref{spec_MPKI_reduction_comparisons} show speedup and MPKI reduction of ACIC, GHRP, 36KB L1i, and OPT over FDP baseline for SPEC2017 Integer Speedup benchmarks with L1i MPKI$>$1. These workloads have high i-cache hit rates even in the baseline, leaving little headroom for ACIC. Still, ACIC does as well as having the larger 36KB L1i.

\subsubsection{With Entangling Prefetching baseline}\label{EIP_evaluation}
\begin{figure}[h!]
\begin{center}
  \includegraphics[width=\linewidth]{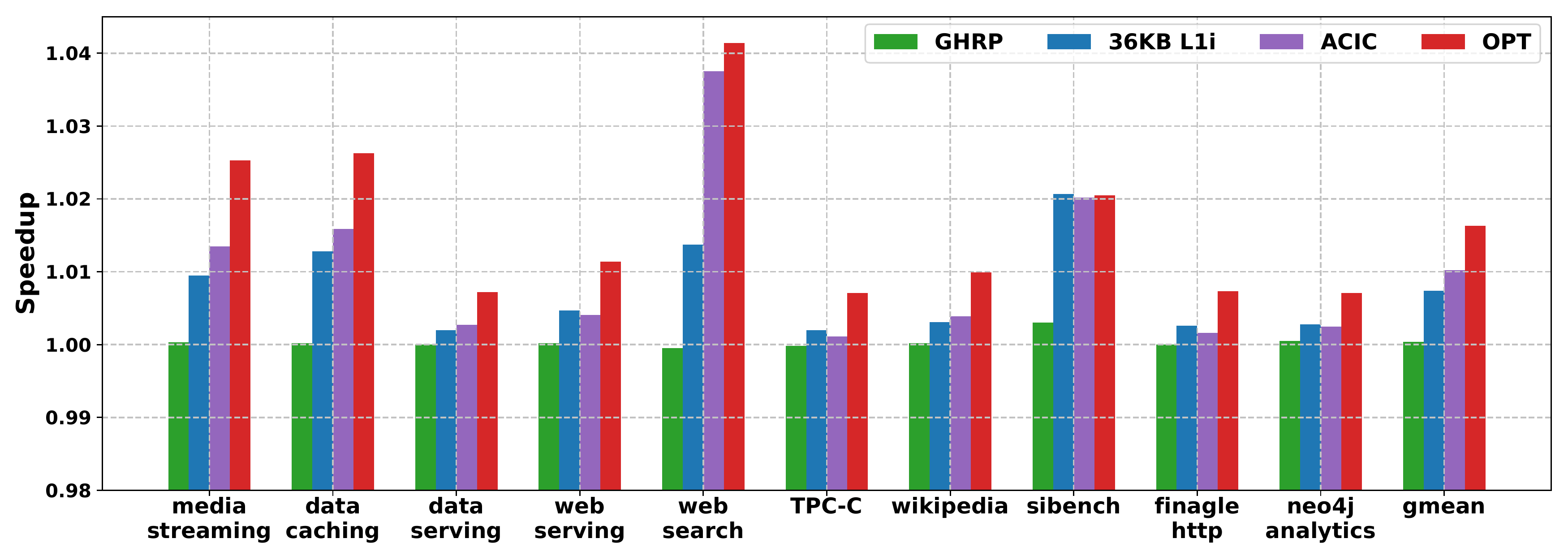}
  \caption{Speedup comparison of various policies over entangling prefetching baseline}
  \label{speedup_comparison_over_EIP}    
\end{center}
\end{figure}

\begin{figure}[h!]
\begin{center}
  \includegraphics[width=\linewidth]{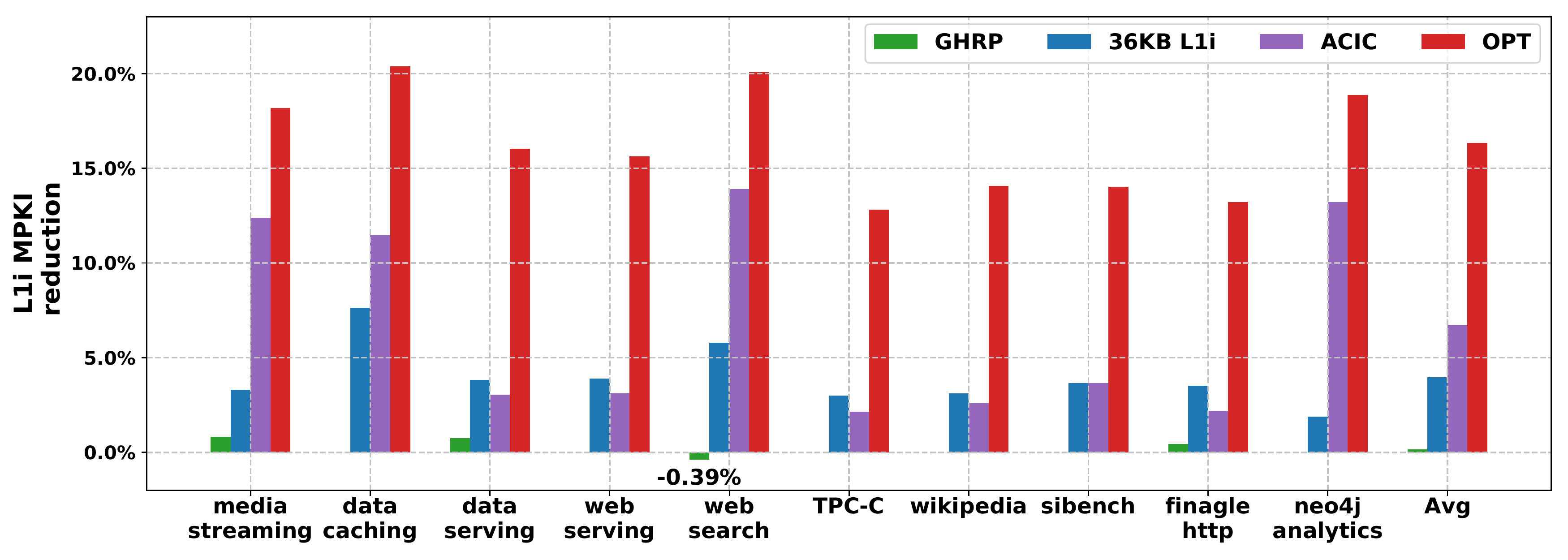}
  \caption{MPKI reduction comparison of various policies over entangling prefetching baseline}
  \label{MPKI_reduction_comparison_over_EIP}    
\end{center}
\end{figure}
Entangling prefetcher~\cite{EIP} is a more recent state-of-the-art instruction prefetcher than FDP. From Figure~\ref{speedup_comparison_over_EIP} and Figure~\ref{MPKI_reduction_comparison_over_EIP}, we can see that with the entangling prefetcher (with a 4K-entry entangled table) baseline, ACIC still outperforms GHRP and 36KB L1i, which are the two best prior policies shown in Figure~\ref{speedup_comparisons} and Figure~\ref{MPKI_reduction_comparisons}. ACIC provides 1.0102 geomean speedup and 6.71\% MPKI reduction over the baseline. Entangling prefetcher improves the baseline L1i hit rate to be over 97\% in our datacenter workloads, so it further complements ACIC's benefits. However, considering that the entangling prefetcher incurs about 40KB storage overhead, which is larger than i-cache itself, ACIC is not redundant. As stated in Section~\ref{motivation}, ACIC and prefetching are complementary, and ACIC can improve i-cache performance beyond the benefits of prefetchers.

\color{black}
\section{Related Work}
The cache pollution problem that ACIC addresses is most closely related to 3 broad categories - replacement policies, bypassing mechanisms and victim caches - that have similar goals, though most of the prior work in these have targetted d-caches as opposed to i-caches.

\noindent {\em Replacement policies:}
There has been considerable work on replacement policies~\cite{software-managed-cache, SRRIP, SDBP, pseudo-LIFO, tree-based-pseudo-LRU, compiler-replacement, LRFU, LRU-k-page, V-way, frequency-based-replacement, EELRU, modified-LRU, counter-based-replacement, cache-bursts, evicted-address-filter}. Since OPT is not implementable, heuristics include variations of LRU~\cite{tree-based-pseudo-LRU, LRU-k-page, EELRU, modified-LRU, LRU-LFU}, frequency~\cite{frequency-based-replacement, LRFU}, reuse prediction~\cite{counter-based-replacement, GHRP, SRRIP, SHiP, SDBP, cache-bursts, dynamic-reuse-distance, Leeway}, and others~\cite{pseudo-LIFO, MRU-insertion, MLP-aware, evicted-address-filter, inter-reference-gap-replacement}. There have also been learning-based policies based on machine learning~\cite{perceptron-learning, multiperspective-reuse-prediction, deep-learning-replacement} and Belady's optimal solution~\cite{Hawkeye, Harmony, Parrot}.

However, as our quantitative evaluation shows, many of these prior proposals for d-caches
(e.g. \cite{SRRIP,SHiP,Hawkeye,Harmony}) do not work as well for i-caches, compared to ACIC. On the other hand, recent techniques for i-caches such as Ripple \cite{Ripple} and GHRP \cite{GHRP} do not identify and leverage the burstiness of accesses to instruction blocks, making ACIC a better alternative as our evaluations have shown.

\noindent {\em Bypassing policies:}
Cache bypassing policies use static approaches~\cite{selective-bypass, compiler-managed-bypassing} with a profile-guided compiler to identify lines for bypassing, and dynamic approaches\cite{runtime_cache_bypassing, reuse-info-bypassing, counter-based-replacement, run-time-adaptive-reference-analysis, dynamic-reuse-distance, multiperspective-reuse-prediction, LLC-block-bypass, hierarchy-aware-bypass, bypass-insertion-LLC, adaptive-bypass-inclusive, DSB, less-reused-filter} which use run-time behavior to learn and predict bypassing opportunities. 

DSB~\cite{DSB} and OBM~\cite{OBM} are two bypassing policies most similar to ACIC in that they also track the reuse behavior of newly allocated cache lines and their corresponding cache contender blocks to learn whether an incoming block should bypass the cache. DSB randomly bypasses newly allocated lines and the effectiveness of the past bypassing decisions is used to tune the bypassing probability. However, unlike the CSHR in ACIC, DSB only tracks one pair in a cache set at a time, and OBM tracks incoming-victim pairs with a low probability to reduce storage overhead. The selective tracking used by DSB and OBM turns out to be much less effective than our CSHR design. 
Moreover, DSB and OBM are further undermined since they are direct bypassing schemes without first separating spatial and temporal locality.

An early work~\cite{runtime_cache_bypassing} uses a small buffer, similar to ACIC, for short temporal/spatial locality. However, they use access counters to compare the utility of incoming and contender blocks, which does not work very well for the instruction stream that exhibits burstiness requiring a more extensive predictor as in ACIC.

\noindent {\em Victim caches:}
Rather than regulate entry, an alternative is to retain the evicted victims temporarily in a victim cache to reduce pollution. 
Works on victim caches~\cite{victim-cache} include ~\cite{Scavenger, identify-conflict-miss, time-keeping}. VVC~\cite{VVC} is a more recent work that predicts dead blocks and reuse the dead regions in the cache as a virtual victim cache. We have shown that ACIC can provide better performance for the instruction stream.

\section{Concluding Remarks \& Future Work}
We drew insight from the observation of bursty accesses in data stream~\cite{cache-bursts}, leveraged i-Filter proposed in~\cite{filter-cache, small-icache} to optimize ``burstiness'' in the instruction stream, and presented an admission control mechanism, ACIC, that regulates the entry of instruction blocks into the i-cache. Comparing with several (8 in all) prior approaches - replacement algorithms, bypassing mechanisms and victim caches - we have shown the benefits of ACIC over these prior approaches. We have also shown that it can complement previously proposed prefetching mechanisms.

The predictor in ACIC learns from on-demand instruction accesses, and tries to estimate reuse distances to implement a more practical version of  Belady's OPT algorithm (comparing reuse distances of i-Filter and i-cache victim). Prefetching could further reduce on-demand misses, but comes at a possible cost of higher memory traffic. As was pointed out in ~\cite{Harmony}, Belady's OPT may not be the best when prefetching is considered. Developing a prefetching-aware ACIC mechanism is part of our future work.


\bibliographystyle{IEEEtranS}
\bibliography{refs}

\end{document}